\documentclass[twocolumn,showpacs,preprintnumbers,amsmath,amssymb]{revtex4}

\usepackage{graphics} \usepackage{dcolumn} \usepackage{bm}

\input epsf
\usepackage{amsmath,amssymb,epsfig,placeins,subfigure,wrapfig,indentfirst,tabularx}

\begin{document}

\title{Homogeneous singularities inside collapsing wormholes}

\author{Elena I. Novikova}
\affiliation{Naval Research Laboratory, Space Science Division,
  Washington, DC}
\author{Igor D. Novikov}
\affiliation{Astro Space Center, Lebedev Physical
  Institute, Russian Academy of Sciences, Moscow, Russia}
\affiliation{The Niels Bohr International Academy, Niels Bohr Institute, Copenhagen,
  Denmark}

\date{\today}

\begin{abstract}
We analyze analytically and numerically the origin of the singularity in the 
course of the collapse of a wormhole  with the exotic scalar field $\Psi$ with 
negative energy density, and with this field $\Psi$ together with the ordered 
magnetic field $H$. We do this under the simplifying assumptions of the spherical 
symmetry and that in the vicinity of the singularity the solution of the 
Einstein equations depends only on one coordinate (the homogeneous approximation). 
In the framework of these assumptions we found the principal difference between 
the case of the collapse of the ordinary scalar field $\Phi$ with the positive 
energy density together with an ordered magnetic field $H$ and the collapse of 
the exotic scalar field $\Psi$ together with the magnetic field $H$. The later 
case is important for the possible astrophysical manifestation of the wormholes.

\end{abstract}


\maketitle

\section{\label{sec:intro}Introduction}

Wormholes (WHs) are hypothetical short topological tunnels
connecting two  different distant asymptotically flat regions
of the Universe, or such regions belonging to different universes in
the model of the Multiverse \cite{Carr07}. They are typical relativistic
objects.

The problem of wormholes has a long history. In the framework of
General Relativity, first attempts to construct such objects were
performed by Flamm \cite{Flamm16} and Einstein and Rozen
\cite{Einstein35}. Later on, many theoretical aspects of the WH
problem were investigated, see, for example \cite{Wheeler55,
  Wheeler57, Misner57, dop-0, Visser95, Ellis73, Kardashev06,
  Kardashev07, Shatskiy07a, Morris88, Novikov89, Frolov98, Thorne93,
  Flanagan96, Bronnikov06, Lemos-Lobo, Armendariz-Picon02, Shatskiy08,
  Doroshkevich08, dop-1, dop-2, dop-3, dop-4, dop-5, Shatskiy07b,
  Dokuchaev04, Dokuchaev05, dop-9, dop-10, dop-11, dop-6, Shinkai02, dop-7,
  dop-8, Nov64-1, Nov64-2, Nov01}. 
During the last few decades, interest in wormholes increased in
connection with the early Universe \cite{Vilenkin83, Linde86,
  Hawking2000, Lobo05} and various problems in physics and
astrophysics \cite{dop-0, Hawking92, Aref07, Shatskiy04a, Shatskiy04b,
  Lobo08, Cherepashchuk05, Hayward09, Hong08, Yeom08, Gonzalez09}.

  In the works \cite{Kardashev06, Kardashev07, Shatskiy07a} the hypothesis that some known astrophysical objects 
  (e.g. quasars and active nuclei of some galaxies) could be entrances 
  to wormholes was considered. Wormholes may have existed as primordial objects 
  in early stage of the expanding Universe \cite{Visser95,  Hawking2000}. It is possible that such 
  primordial wormholes could be preserved after the end of the
  inflation \cite{Lobo05, Hayward09}. 
  This hypothesis can explain some observable facts in astrophysics and can predict 
new phenomena \cite{Kardashev07}.
  
  In the middle of the last century it was shown in General Relativity that a vacuum 
  WH pinches so quickly that it cannot be traversed even by a test signal 
  moving with the velocity of light (see review in \cite{Frolov98}). In order to prevent the 
  shrinking of a WH and to make it traversable, it is necessary to thread its throat
  with so-called exotic matter which is matter that violates the averaged null energy 
  conditions (see \cite{Visser95, Frolov98, Thorne93, Flanagan96}). 
  Different types of wormholes may exist depending on 
  the type of exotic matter in their throats \cite{Kardashev07,
  Armendariz-Picon02, Shatskiy08}. 
  For example it could be a
  ``magnetic exotic matter'' in which the main component is a strong ordered magnetic 
  field plus a some amount of a ``true exotic matter'' \cite{Kardashev07}. Another type is a ``scalar  
  exotic matter'' in the form of a scalar field with a negative energy density \cite{Ellis73, Armendariz-Picon02}.  
  One more type is a mixture; ``magnetic-negative dust exotic matter'' which is a 
  mixture of an ordered magnetic field and dust (matter with zero pressure) with 
  negative matter density \cite{Shatskiy08}.  The physical properties of different types of WHs 
  are different. WHs with scalar exotic matter were the subject  of very intense 
  investigations both analytically and numerically. Possible dynamics of such WHs has 
  been analyzed analytically for example in papers 
  \cite{Doroshkevich08, dop-1, dop-2, dop-3, dop-4, dop-5, Shatskiy07b,
  Dokuchaev04, Dokuchaev05}. The most important, however, was 
  the numerical analysis \cite{dop-6, Shinkai02, dop-7, Doroshkevich09}. It was
  shown that the static WHs of this type are unstable.  Perturbations trigger the evolution of the WH and the evolution of the 
  exotic scalar field which maintains it. As a result, the WH either collapses or 
  expands. In the case of the collapse - a black hole (BH) arises. 
  If the collapsing WH has both the exotic scalar
  field and the magnetic field, the structure of the singularity
  inside the resulting BH requires special investigation.
  The reason for extra attention is that the magnetic 
  field in WHs or their remnants has special manifestation in the astrophysical 
  observations in the hypothesis of the existence of WHs in the Universe  \cite{Kardashev07}.
  
  The goal of this paper is to analyze (under simplifying assumptions) the nature of the 
  singularity created during the collapse of the WH with the exotic scalar field and with 
  the magnetic field.
  
  The paper is organized as follows.
  In section \ref{sec:model} we describe a model in the framework of which we analyze the problem.
  In the subsection \ref{sec:model_equations} the equations are written out. In the subsection \ref{sec:model_leadOrderq=0} we use the
  leading order analysis for the case of a WH without  magnetic field. In the subsection
  \ref{sec:model_numericalAnalysis} we use a numerical code to solve the equations to understand the behavior of the
  model for several sets of its parameters. In subsection \ref{sec:model_leadOrderqNE0} we use the leading order
  analysis for the case with the magnetic field. In the subsection \ref{sec:model_numericalAnalysisEpsLess0} the numerical 
  analysis is applied to the case with the magnetic field. In the section \ref{sec:conclusions} we discuss 
  the results.
  
    \section{\label{sec:model} The model}
  
 We consider collapse of a WH with the formation of a BH. The main
 feature inside a BH is its space-time singularity. 
 Our goal is to investigate the nature of this singularity, using some
 simplifying assumption. 
 First, we consider the spherical WH and BH nonlinearly perturbed by a minimally coupled
 and self-gravitating massless exotic scalar field $\Psi$ with the negative energy density 
 $\epsilon <0$. We will also consider the same case but with an additional radial 
 magnetic field. It was shown (see \cite {Burko97, Burko98, Hansen05})
 that, in the considered model, in the close vicinity of a 
 space-like singularity of a BH, all processes, as a rule, have high temporal 
 gradient (much higher than the spatial gradients along the
 singularity) and that the processes 
 depend on the properties of a very restricted space region. It
 follows from here that, for clarification of some physical 
 processes, one can use a homogenous approximation and that all processes and geometry 
 depend on the time coordinate only. We assume also that in the close vicinity of a 
 time-like singularity of a BH one can use a homogenous approximation also, 
 but now all processes and geometry depend on the radial space coordinate only.
  
 \subsection{\label{sec:model_equations}The equations}
  
  We start with the general homogeneous spherically symmetric line element:
\begin{equation}
\label{eq:1}
  ds^2=g_{tt}(r)dt^2+g_{rr}(r)dr^2+r^2d\Omega^2
\end{equation}
\begin{equation}
\label{eq:2}
  d\Omega^2=d\Theta^2+\sin^2\Theta d\phi^2
\end{equation}
Inside a BH the region between the event horizon (EH) and the Cauchi horizon 
(which exists in the case with the radial magnetic field; we will call it \hbox{horizon-2} (\hbox{H-2})) 
is so-called $T$-region (see \cite{Frolov98} and references therein). 
In this region $r$ is time-like and $t$ is space-like coordinate. To
describe the contraction, we should consider the variation of the time coordinate $r$ from bigger 
to smaller values.
 The $r-r$, $t-t$ and $\Theta-\Theta$ components of the Einstein 
equations (with $c=1$, $G=1$) are given by (see \cite{Hansen05})

\begin{equation}
  \label{eq:3} 
  \frac{g_{tt}-g_{rr}g_{tt}+rg\prime_{tt}}{r^2g_{rr}g_{tt}}=8\pi(T_r^r+H_r^r)
\end{equation}
\begin{equation}
  \label{eq:4}
  \frac{g_{rr}-g_{rr}^2-rg\prime_{rr}}{r^2g_{rr}^2}=8\pi(T_t^t+H_t^t)
\end{equation}

\begin{eqnarray}
 \label{eq:5}
  \frac{1}{4r g_{rr}^2g_{tt}^2}
          \bigg\{g_{tt}\bigg[2g_{rr}(g\prime_{tt}+rg\prime\prime_{tt})-
      (rg\prime_{rr}g\prime_{tt})\bigg]-\\
\nonumber
      2g_{tt}^2g\prime_{rr}-rg_{rr}g\prime_{tt}^2\bigg\}=
       8\pi\bigg(T_{\Theta}^{\Theta}+H_{\Theta}^{\Theta}\bigg)
\end{eqnarray}
  where the primes denote differentiation with respect to $r$. Tensor $T$ represent 
  here contribution from the $\Psi$ field and tensor $H$ represents contribution from 
  a free radial magnetic field. In the WH, $H$ is a sourceless magnetic field captured 
  by the topological structure of the 3-D geometry. For both the WH and 
  arising from the collapse of it BH the components of the tensor $H$ are given by:
 \begin{equation}
 \label{eq:6}
  H^r_r=H^t_t=-H_{\Theta}^{\Theta}=-\frac{q^2}{8\pi r^4},
\end{equation} 
where the constant $q$ characterizes the strength of the magnetic field. The exotic 
scalar field $\Psi$ is governed by the Klein-Gordon equation $\Psi^{;\alpha}_{;\alpha}=0$  
(; denotes the covariant derivative), whose first integral reads 
\begin{equation}
 \label{eq:7}
  \Psi^\prime(r)=\frac{d}{\sqrt{-g}}g_{rr}\sin{\Theta}
\end{equation}
where $d$ is a constant and $g$ is the metric determinant. For the
exotic field, the value of $d$ is 
pure imaginary, $d^2<0$. Of course $\Phi^\prime$ does not depend on $\Theta$ because 
the metric determinant $g$ has a factor   $\sin^2{\Theta}$.

The components of the tensor $T$ for the exotic scalar filed $\Psi$ are (see \cite{Hansen05}):
\begin{equation}
 \label{eq:8}
 T^r_r=-\epsilon
\end{equation}
\begin{equation}
 \label{eq:9}
 T^t_t= \epsilon 
\end{equation}
\begin{equation}
 \label{eq:10}
 T^{\Theta}_{\Theta}=\frac{\epsilon}{g_{rr}}
\end{equation}
\begin{equation}
 \label{eq:11}
 \epsilon=\epsilon_0\bigg(\frac{g_{tt,init}}{g_{tt}}\bigg)\bigg(\frac{r_{init}}{r}\bigg)^4=
 -\frac{1}{8\pi g_{rr}}(\Psi^\prime)^2
\end{equation}
\begin{equation}
 \label{eq:12}
  \epsilon_0=\frac{d^2}{8\pi\cdot g_{tt,init}\cdot r^4_{init}}
\end{equation}
where $\epsilon_0$, $g_{tt,init}$, $r_{init}$, $d^2<0$ are all 
constants. Substitution of (\ref{eq:6}, \ref{eq:8} -  \ref{eq:12})
 into (\ref{eq:3}), (\ref{eq:4}) enables us to find the unknown functions $g_{rr}(r)$ and 
$g_{tt}(r)$. Equation (\ref{eq:5}) is the consequence of (\ref{eq:3}), (\ref{eq:4}) and hence can be used as a 
control of the calculations.

\subsection{\label{sec:model_leadOrderq=0}Leading order analysis for the case $q=0$}

We will use the method which has been proposed by Burko \cite{Burko97,
  Burko98} and after that used 
in \cite{Hansen05}. This is the non-linear  generalization in the homogenous case of linear 
analysis of \cite{Doroshkevich78}. In this section we consider the
  case \hbox{ $q=0$}. Let us consider the 
leading order terms in a series expansion  for the metric  functions and the leading 
order terms in the Einstein equations  (\ref{eq:3}), (\ref{eq:4}),
  $q=0$ near the singularity $r=0$.
We assume that the leading terms for each $g_{tt}$ and $g_{rr}$ can be
  written in the power form ($const\cdot r^{power}$). We
  demand also that the corresponding terms in the Einstein equations
  (\ref{eq:3}), (\ref{eq:4}) tend to zero when $r\rightarrow 0$,
  otherwise this method doesn't work (see Appendix \ref{append}).
In the vicinity 
of $r=0$ the solution of  (\ref{eq:3}), (\ref{eq:4}) can be written, as
  the first approximation, in the 
following form
\begin{equation}
 \label{eq:13}
  g^{(1)}_{tt}=2m_0Cr^{\beta},
\end{equation}
\begin{equation}
 \label{eq:14}
 g^{(1)}_{rr}=-(\beta+2)\frac{1}{2m_0}r^{\beta+2}, 
\end{equation}
where $m$, $C$ and $\beta$ are all constants.  For the scalar field
$\Psi$, the leading order (in $r$) analysis gives:
\begin{equation}
 \label{eq:15}
 \Psi^{(1)}_{(r)}=\sqrt{\beta+1}\ln{r}. 
\end{equation}
We have also for the constant $d^2$ (see (\ref{eq:7}) and (\ref{eq:12})):
\begin{equation}
 \label{eq:16}
 d^2=\frac{(\beta+1)}{(\beta+2)}\cdot 4m_0^2C. 
\end{equation}
Here we use constants $m_0$ and $C$ which were introduced in \cite{Hansen05}. These constants 
have direct physical meaning in the case of empty (without $\Psi$ field) BH, when 
\hbox{$\beta=-1$}. In such a  case, $m$ is the BH mass, $C$ is a gauge parameter related to 
the possibility of changing the scale of measurement of the $t$ space coordinate.

In the case of presence of the $\Psi$ field, $\beta$ determines the strength of this 
field. In contrast to the case analyzed by Burko in  \cite {Burko97, Burko98}, in our case here 
$\Psi$ should be imaginary and from (\ref{eq:16}) we have the restriction:
\begin{equation}
 \label{eq:17}
 \beta<-1. 
\end{equation}

On the other hand, for $m_0>0$ we obtain from (\ref{eq:13}) and (\ref{eq:14}):
\begin{equation}
 \label{eq:18}
 C>0, 
\end{equation}
\begin{equation}
 \label{eq:19}
 \beta+2>0. 
\end{equation}
Thus we have for $\beta$ the restrictions $-2<\beta<-1$. In the case of violation 
of the inequality (\ref{eq:19}) this method does not work as we mentioned above and in 
Appendix A.

The above solution (\ref{eq:13}), (\ref{eq:14}) is generic, because it depends on three arbitrary parameters. 
In our case $m$ and $\beta$ are essential physical parameters, and $C$ is a gauge 
parameter. Using the arguments which are analogous to the arguments of Burko \cite {Burko97, Burko98} 
 one may conclude that the dependence on two physical and one gauge parameter 
means that the solution is generic. If we suppose that in the solution (\ref{eq:13}), (\ref{eq:14}) $m_0<0$, 
$C>0$, then it means that $g_{tt}^{(1)}<0$, $g_{rr}^{(1)}>0$; $t$ is the time coordinate 
and $r$ is the space coordinate. In this case $r=0$ is a time-like singularity. Probably 
it corresponds to negative mass of the object and a naked singularity.

\subsection{\label{sec:model_numericalAnalysis}Numerical analysis}

We use a simple numerical code to solve numerically equations  (\ref{eq:3}), (\ref{eq:4}) to understand 
the behavior of model (\ref{eq:1}), (\ref{eq:2}) as a function of the parameters of the model.

We will represent the results for the case of the exotic scalar field $\Psi$ with
$\epsilon<0$, and compare them with the results for the case of the ordinary scalar 
field $\Phi$ with $\epsilon>0$.  For the latter case $\Phi$-field with $\epsilon>0$, we 
consider also the case of the presence of the magnetic field $q\neq 0$ to emphasize the 
essential difference the cases $q=0$ and $q\ne 0$.

We start the computation from $r_{init}=0.95r_{EH}$, and set  $q_{tt,init}$, $q_{rr,init}$
equal to their values at $r_{init}$ for the zero matter content Schwarzschild 
(or Reisner-Nordstr\"{o}m if $q\neq 0$) solution with initial $m_0=1$ and $q=0.95$ if 
$q\neq 0$. 
Here $r_{EH}$ is the value of the event horizon for the zero matter content BH 
with the same parameters. We consider several values of
$\epsilon_0$, which is the characteristic of the initial 
amplitude of the scalar field.

\begin{figure*}
\subfigure[\hskip 1mm  $r$ versus $t$. \label{fig:1a}]{\includegraphics[width=0.495\textwidth]{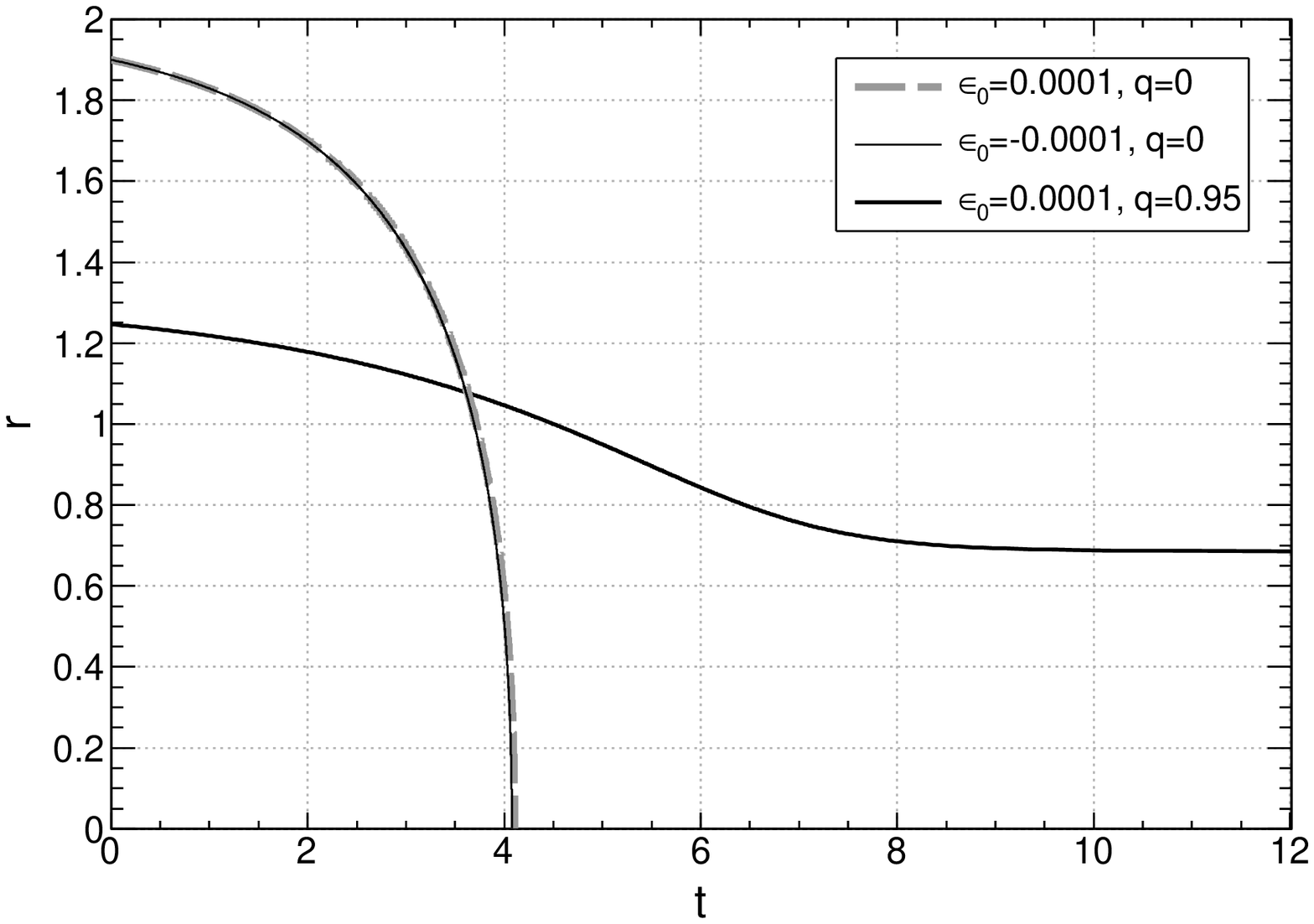}}
\subfigure[\hskip 1mm  Mass function versus $r$. \label{fig:1b}]{\includegraphics[width=0.495\textwidth]{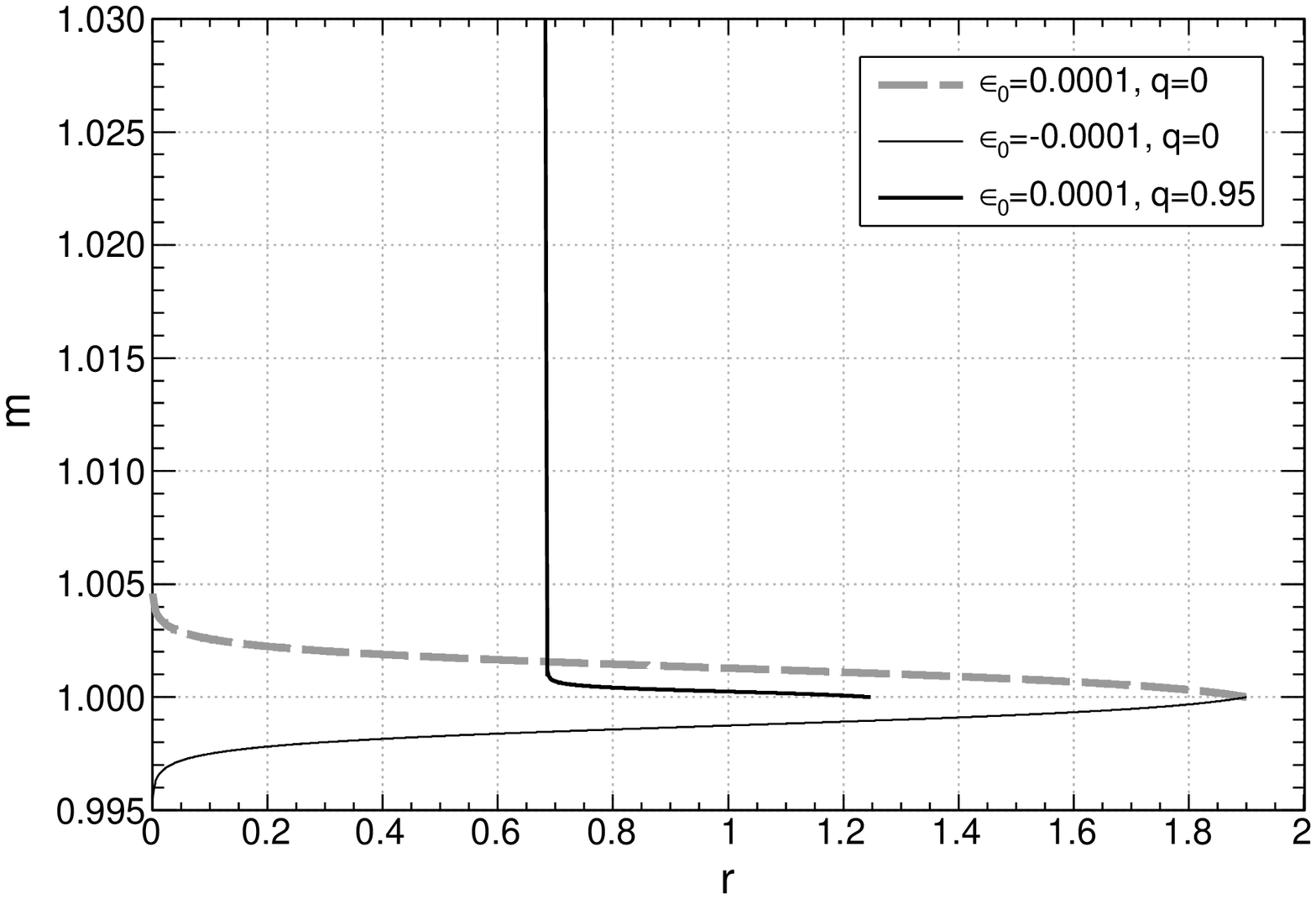}}
\subfigure[\hskip 1mm  Metric functions $g_{tt}$ and  $|g_{rr}|$ vs $r$. \label{fig:1c}]{\includegraphics[width=0.495\textwidth]{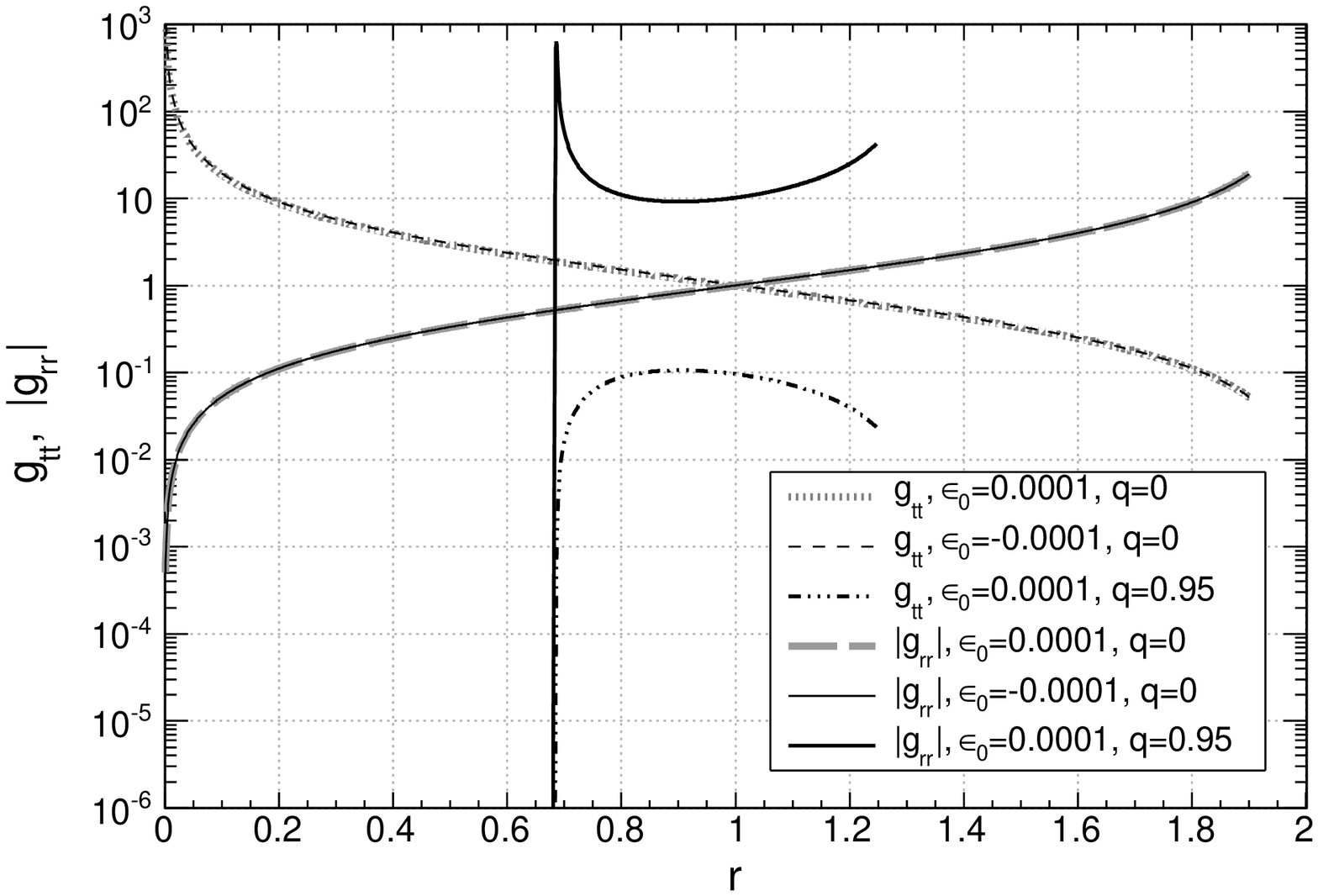}} 
\subfigure[\hskip 1mm  The $d\log g_{tt}/d\log r$ \hskip 1mm and\hskip 2mm
  $d\log|g_{rr}|/d\log r$ \hskip 2mm   vs \hskip 2mm
  $r$. \label{fig:1d}]{\includegraphics[width=0.495\textwidth]{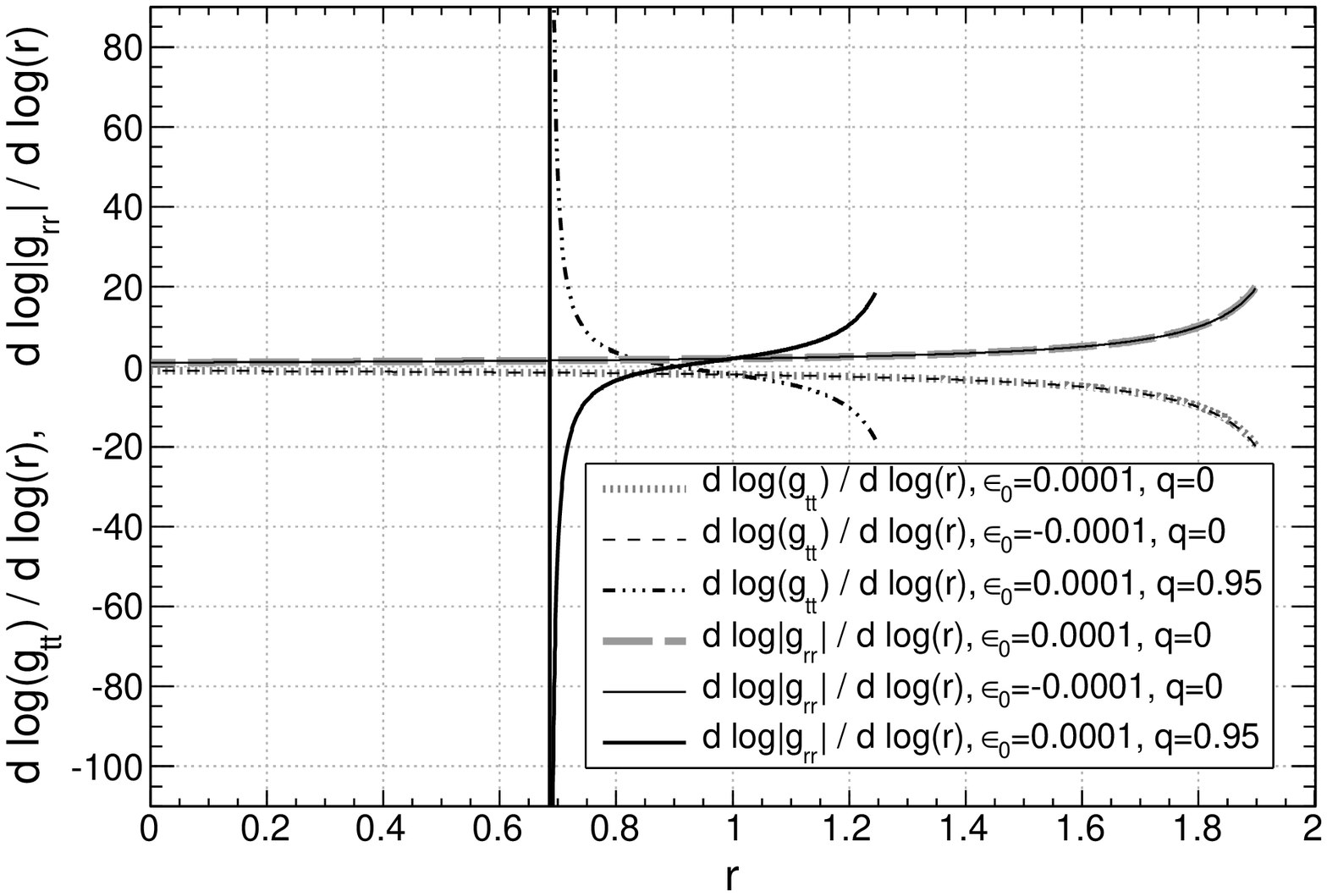}}
\caption{\label{fig:1} The cases of $\epsilon_0=0.0001, q=0$, \hskip 2mm
    $\epsilon_0=-0.0001, q=0$, \hskip 2mm and  $\epsilon_0=0.0001,
    q=0.95$.}
\end{figure*} 

The parameters of the model, including the exponent $\beta$, depend on the value of 
the scalar field $\epsilon_0$. We start from the extremely small amplitude 
$|\epsilon_0|$ and consider the cases (see Figure \ref{fig:1}):
\begin{eqnarray}
 \label{eq:20}
 \epsilon_0=+0.0001, & q=0, \\
 \label{eq:21}
 \epsilon_0=-0.0001, & q=0, \\
 \label{eq:22}
 \epsilon_0=+0.0001, & q=0.95, 
\end{eqnarray}
In Figure \ref{fig:1a} one can see the propagation ($r$ vs. $t$) of the incoming signal 
with the velocity $c$ ($c=1$).  Figure \ref{fig:1b} shows the mass function (see \cite{Hansen05}):
\begin{equation}
 \label{eq:23}
 m=\frac{r}{2}\bigg(1+\frac{d^2}{r^2}-g^{-1}_{rr}\bigg). 
\end{equation}
Figure \ref{fig:1c} presents the evolution of the metric functions, $g_{tt}$ and $g_{rr}$, and Figure \ref{fig:1d}
shows the evolution of the values $d\log g_{tt}/d\log r\equiv\beta$
and $d\log |g_{rr}|/d\log r\equiv\alpha$.

It is clear from the results shown in Figures \hbox{\ref{fig:1c}--\ref{fig:1d}} that for this extremely small amplitude 
$|\epsilon_0|$ in the case $q=0$ the propagation of the light-like signal, evolution of 
$g_{tt}$ and $g_{rr}$ and \hbox{$d\log g_{tt}/d\log r=\beta$}  and \hbox{$d\log
|g_{rr}|/d\log r=\alpha$} 
are practically the same for \hbox{$\epsilon_0=\pm 0.0001$} and the difference is not visible 
in the plots.

On the other hand, the behavior of the mass function for the cases
$\epsilon_0=0.0001$  and $\epsilon_0=-0.0001$ (Figure \ref{fig:1a}) is quite different. Of course 
the reason for this is the opposite signs of $\epsilon$ for these two
cases. 

The physical processes which can lead to a nonlinear change of the mass function are 
the following:

\begin{itemize}
\item[{\bf A)}] 
The mass $m$ inside a sphere can change because of the work of pressure forces on 
the surface of the sphere,
\item[{\bf B)}] 
The mass can change due to the mass inflation  \cite{Hansen05}.
\end{itemize}

Both processes are described in  \cite{Hansen05}. The process {\bf B)} is important near the \hbox{horizon-2} \hbox{$(H\mbox{-}2)$} 
in the case of presence of the magnetic field $q\neq 0$ and $\epsilon >0$. 
In the case $q=0$ \hbox{horizon-2} \hbox{$(H\mbox{-}2)$} does not exist.
Thus in the case $q=0$ the process {\bf B)} is not important, and variations of mass 
function is small and opposite for the two cases $\epsilon_0=0.0001$  and $\epsilon_0=-0.0001$. 

The case of 
$\epsilon_0=0.0001$, $q=0.95$ is quite different. Figure \ref{fig:1a} shows that in this case the 
light-like signal goes close to the initial value of $r$ for the
\hbox{$(H\mbox{-}2)$} horizon 
$r=r_{H\mbox{-}2}$ during a long period, and only after that it comes to $r=0$ (beyond the 
framework of the figure).
Note that the light-like signal does not cross the $H\mbox{-}2$ horizon. The horizon itself 
shrinks down to $r=0$ under the focusing effect, as it is described in \cite{Hansen05}.
The metric functions $g_{tt}$ and $g_{rr}$ go to zero very fast, and very large asymptotic 
values $d\log g_{tt}/d\log r=\beta$ and  $d\log |g_{rr}|/d\log r=\alpha$ are clearly seen in Figure \ref{fig:1d}. 
When the solution is close to $r=r_{H\mbox{-}2}$ the mass inflation
manifests itself (see Figure \ref{fig:1b}),
the mass function increases dramatically.

\begin{figure*}
\subfigure[\hskip 1mm  $r$ versus $t$. \label{fig:2a}]{\includegraphics[width=0.495\textwidth]{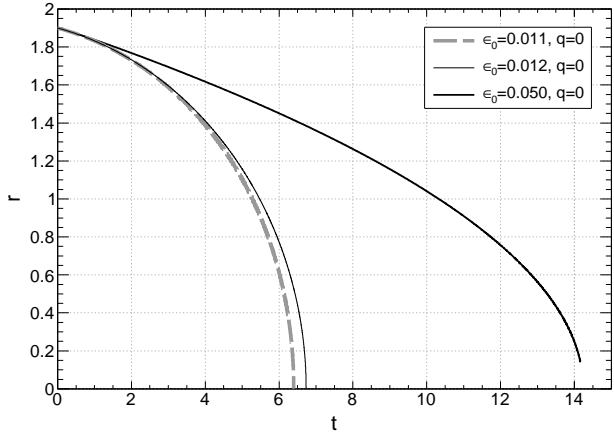}}
\subfigure[\hskip 1mm  Metric functions $g_{tt}$ and  $|g_{rr}|$  vs  $r$. \label{fig:2b}]{\includegraphics[width=0.495\textwidth]{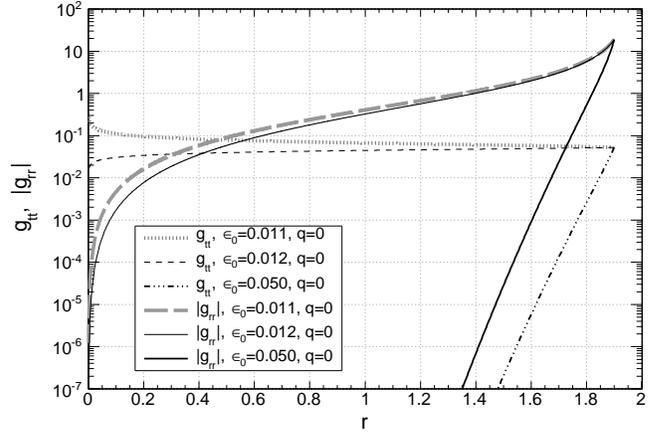}} 
\subfigure[\hskip 1mm  The $d\log g_{tt}/d\log r$ \hskip 1mm and\hskip 2mm  $d\log|g_{rr}|/d\log r$ \hskip 2mm   vs \hskip 2mm  $r$.
  \label{fig:2c}]{\includegraphics[width=0.495\textwidth]{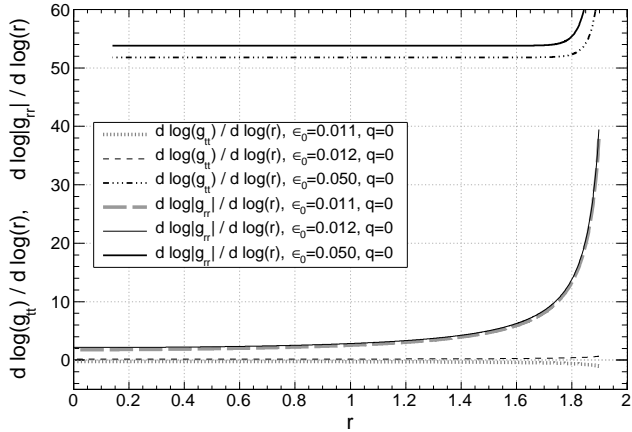}}
\subfigure[\hskip 1mm  Mass function versus $r$. \label{fig:2d}]{\includegraphics[width=0.495\textwidth]{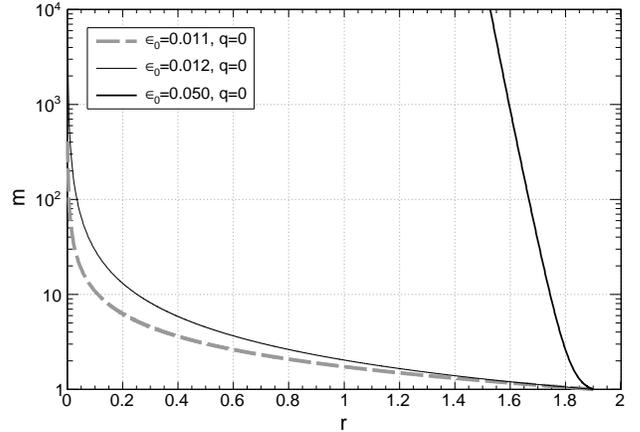}}
\caption{\label{fig:2} The cases of $\epsilon_0=0.011, q=0$, \hskip 2mm
    $\epsilon_0=0.012, q=0$, \hskip 2mm and  $\epsilon_0=0.050,
    q=0$.}
\end{figure*} 

We next study the solutions for the larger values of $|\epsilon_0|$.  We start with
the case $\epsilon_0>0$ and consider the following values (see Figure \ref{fig:2}):
\begin{eqnarray}
 \label{eq:24}
\epsilon_0=0.011, & q=0,\\
 \label{eq:25}
\epsilon_0=0.012, & q=0,\\
 \label{eq:26}
\epsilon_0=0.050,  & q=0,
\end{eqnarray}

The propagation of the light-like signal (Figure \ref{fig:2a}) qualitatively is the same for all 
cases but for larger values of $\epsilon_0$ it comes to $r=0$ later.
behavior of the metric functions, Figures \ref{fig:2b}--\ref{fig:2c}, is interesting. The analysis of the leading 
order terms of the series expansion for the case $\epsilon>0$ (this means that $d^2>0$) 
leads to the conclusion that $\beta > -1$, but $\beta$ can be positive and negative. 
Numerical analysis  shows that for rather small $\epsilon_0\leq 0.011$ the function 
$g_{tt}$ 
increases when $r\rightarrow 0$, but for the larger $\epsilon_0\geq 0.012$ the function 
$|g_{tt}|$ decreases when $r\rightarrow 0$ (see also Table \ref{table:1}).

\begin{table}
 \caption{\label{table:1} The limit values of  $(d\log g_{tt}/d\log
 r)\equiv\beta$ and $(d\log |g_{rr}|/d\log r)\equiv\alpha$\hskip 2mm for
 \hskip 2mm $q=0$,\hskip 3mm for \hskip 2mm $r\rightarrow 0$.}
  \begin{tabularx}{4.5cm}{X X X}
    \hline
       \parbox[c]{1.5cm}{$\epsilon_0$} & \parbox[c]{1.5cm}{$\alpha$} & \parbox[c]{1.5cm}{$\beta$} \\
     \hline
       \parbox[c]{1.5cm}{0.0001} & \parbox[c]{1.5cm}{1.00} & \parbox[c]{1.5cm}{-1.00} \\
       \parbox[c]{1.5cm}{0.0010} & \parbox[c]{1.5cm}{1.01} & \parbox[c]{1.5cm}{-0.99} \\
       \parbox[c]{1.5cm}{0.0025} & \parbox[c]{1.5cm}{1.02} & \parbox[c]{1.5cm}{-0.98} \\
       \parbox[c]{1.5cm}{0.0005} & \parbox[c]{1.5cm}{1.06} & \parbox[c]{1.5cm}{-0.94} \\
       \parbox[c]{1.5cm}{0.0100} & \parbox[c]{1.5cm}{1.52} & \parbox[c]{1.5cm}{-0.48} \\
       \parbox[c]{1.5cm}{0.0110} & \parbox[c]{1.5cm}{1.79} & \parbox[c]{1.5cm}{-0.21} \\
       \parbox[c]{1.5cm}{0.0120} & \parbox[c]{1.5cm}{2.16} & \parbox[c]{1.5cm}{+0.16} \\
       \parbox[c]{1.5cm}{0.0200} & \parbox[c]{1.5cm}{8.95} & \parbox[c]{1.5cm}{+6.95} \\
       \parbox[c]{1.5cm}{0.0500} & \parbox[c]{1.5cm}{53.8} & \parbox[c]{1.5cm}{+51.8} \\
    \hline    
  \end{tabularx}
\end{table}

As it can be seen from Table \ref{table:1} the function $\lim _{r\rightarrow 0}{\beta}$ where 
$\beta\equiv d\log g_{tt}/d\log r$ increases monotonically when $\epsilon_0$ varies 
from $\beta=-1$ when $\epsilon_0=0$, to $\beta=+51.8$ when $\epsilon_0=0.05$. 
The critical value of $\epsilon_0$, when $\beta$ changes its sign, is 
$\epsilon_{0,crit}\approx 0.0115$.  The value $\lim_{r\rightarrow 0}{\alpha}$ where
$\alpha\equiv d\log |g_{rr}|/d\log r$ also increases monotonically when $\epsilon_0$ 
varies from $\alpha=1$ when $\epsilon_0 =0$ to $\alpha =53.8$ when $\epsilon_0=0.05$. 
The sign of $\alpha$ never changes.  

The difference in behavior of the metric functions leads
to difference in the properties of the mass function.
As we mentioned above, in the case $q=0$ the horizon
\hbox{$H\mbox{-}2$} does not exist and the mass inflation (the process {\bf B)})
is not important for the evolution of the mass function. Thus, the
process {\bf A)}, which is related to the deformation of 
the volume of the reference frame, plays the main role. The longitude  deformation is proportional to 
$(g_{tt})^{1/2}$, while the transversal deformations are proportional to $r^2$. In all cases it 
leads to the increase of the mass with  decrease of $r$.  For large positive $\beta$ 
corresponding to large $\epsilon_0$ this increase is very fast.

\begin{figure*}
\subfigure[\hskip 1mm  Metric functions $g_{tt}$ and  $|g_{rr}|$  vs
  $r$. \label{fig:3a}]{\includegraphics[width=0.495\textwidth]{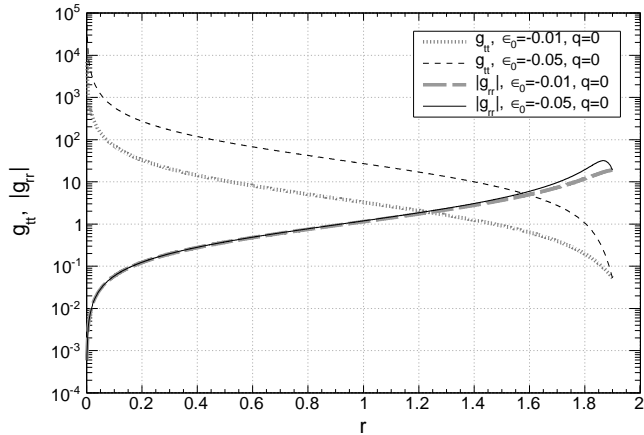}} 
\subfigure[\hskip 1mm  The $d\log g_{tt}/d\log r$ \hskip 1mm and\hskip 2mm  $d\log|g_{rr}|/d\log r$ \hskip 2mm   vs \hskip 2mm  $r$.
  \label{fig:3b}]{\includegraphics[width=0.495\textwidth]{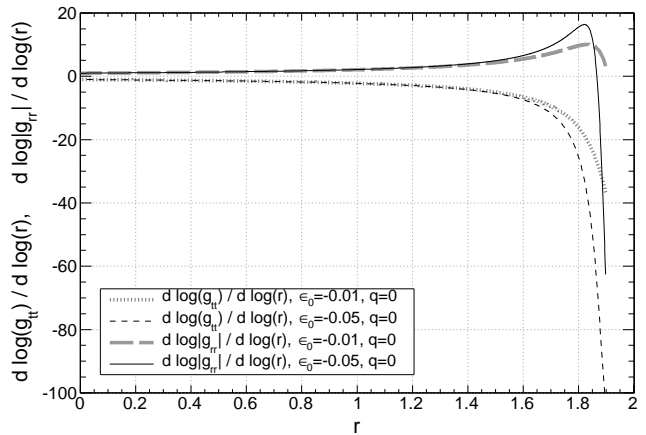}}
\subfigure[\hskip 1mm  $r$ versus $t$. \label{fig:3c}]{\includegraphics[width=0.495\textwidth]{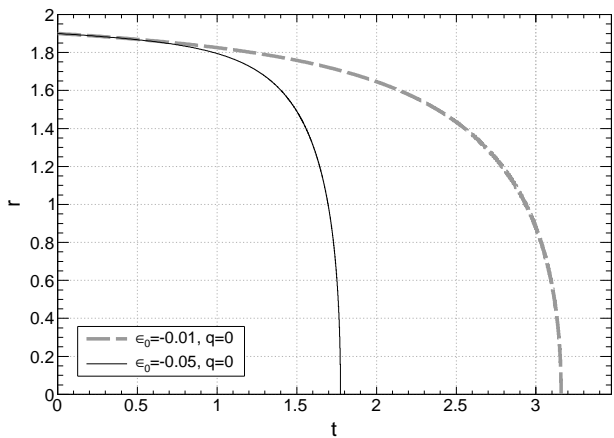}}
\subfigure[\hskip 1mm  Mass function versus $r$. \label{fig:3d}]{\includegraphics[width=0.495\textwidth]{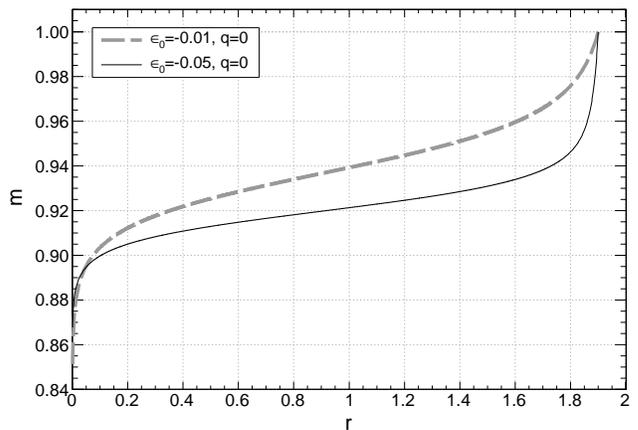}}
\caption{\label{fig:3} The cases of $\epsilon_0=-0.01, q=0$, \hskip 2mm  and  $\epsilon_0=-0.05, q=0$.}
\end{figure*} 

\begin{figure*}
\subfigure[\hskip 1mm  Metric functions $g_{tt}$ and  $|g_{rr}|$  vs
  $r$. \label{fig:4a}]{\includegraphics[width=0.495\textwidth]{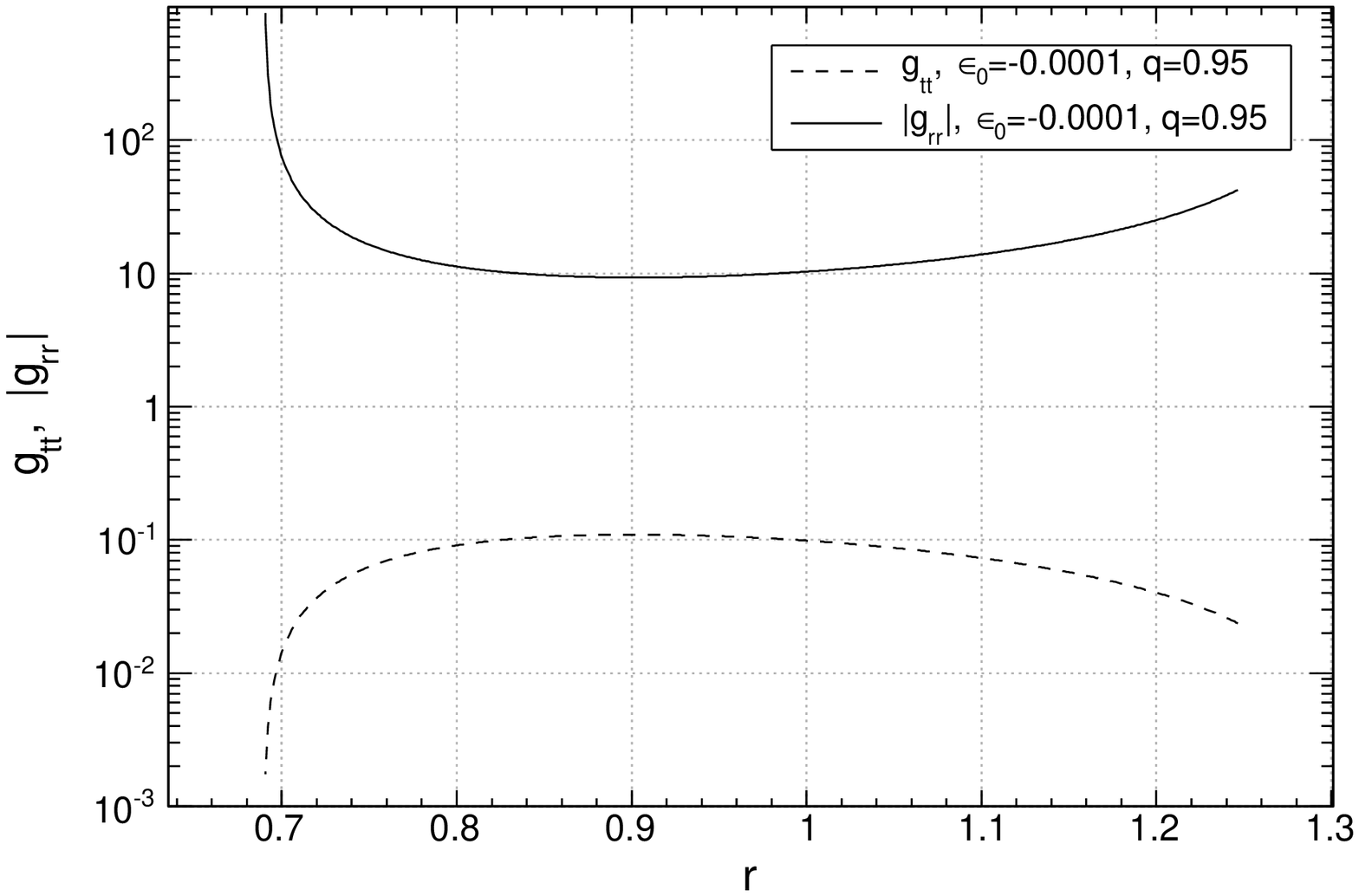}} 
\subfigure[\hskip 1mm  The $d\log g_{tt}/d\log r$ \hskip 1mm and\hskip 2mm  $d\log|g_{rr}|/d\log r$ \hskip 2mm   vs \hskip 2mm  $r$.
  \label{fig:4b}]{\includegraphics[width=0.495\textwidth]{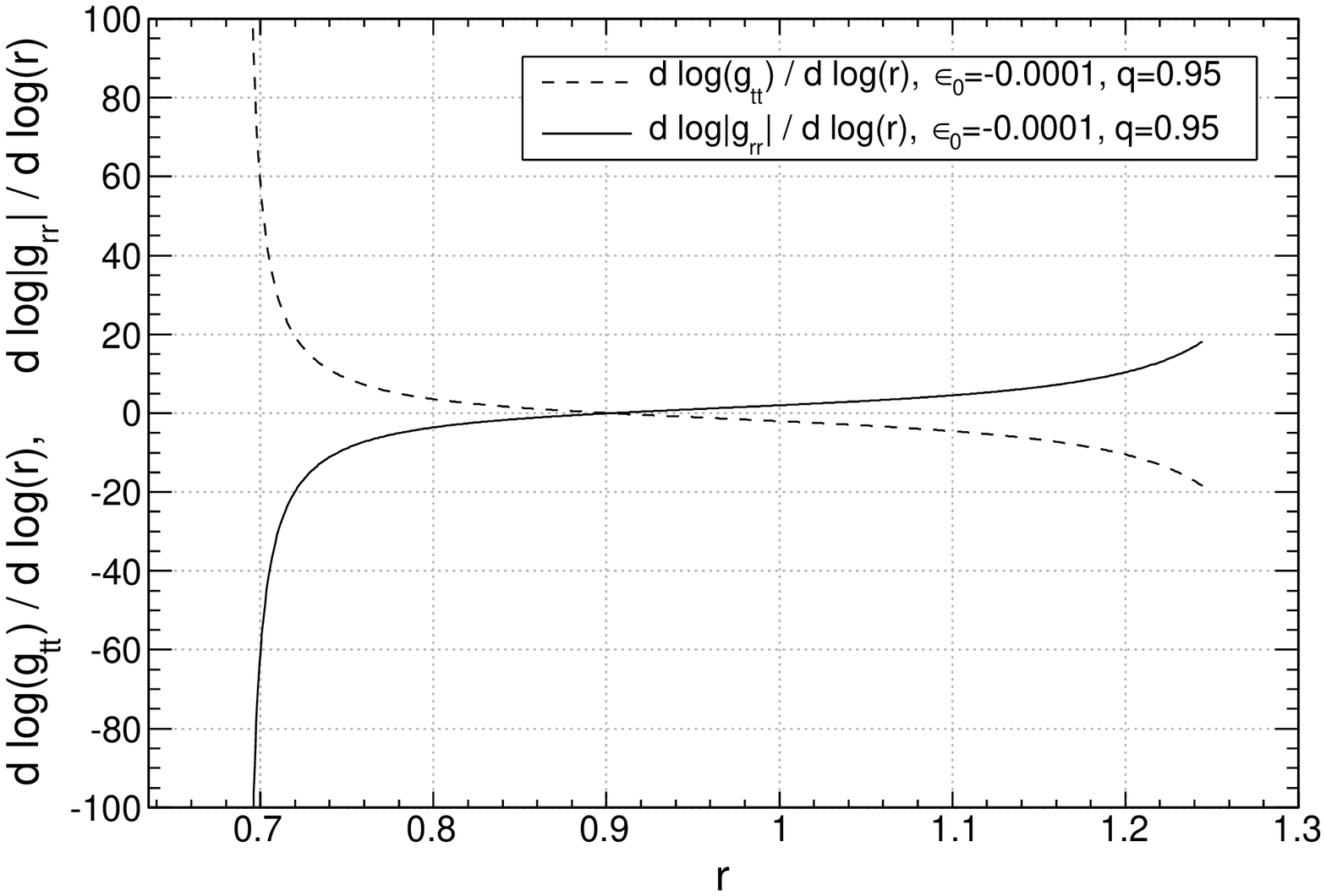}}
\subfigure[\hskip 1mm  $r$ versus $t$. \label{fig:4c}]{\includegraphics[width=0.495\textwidth]{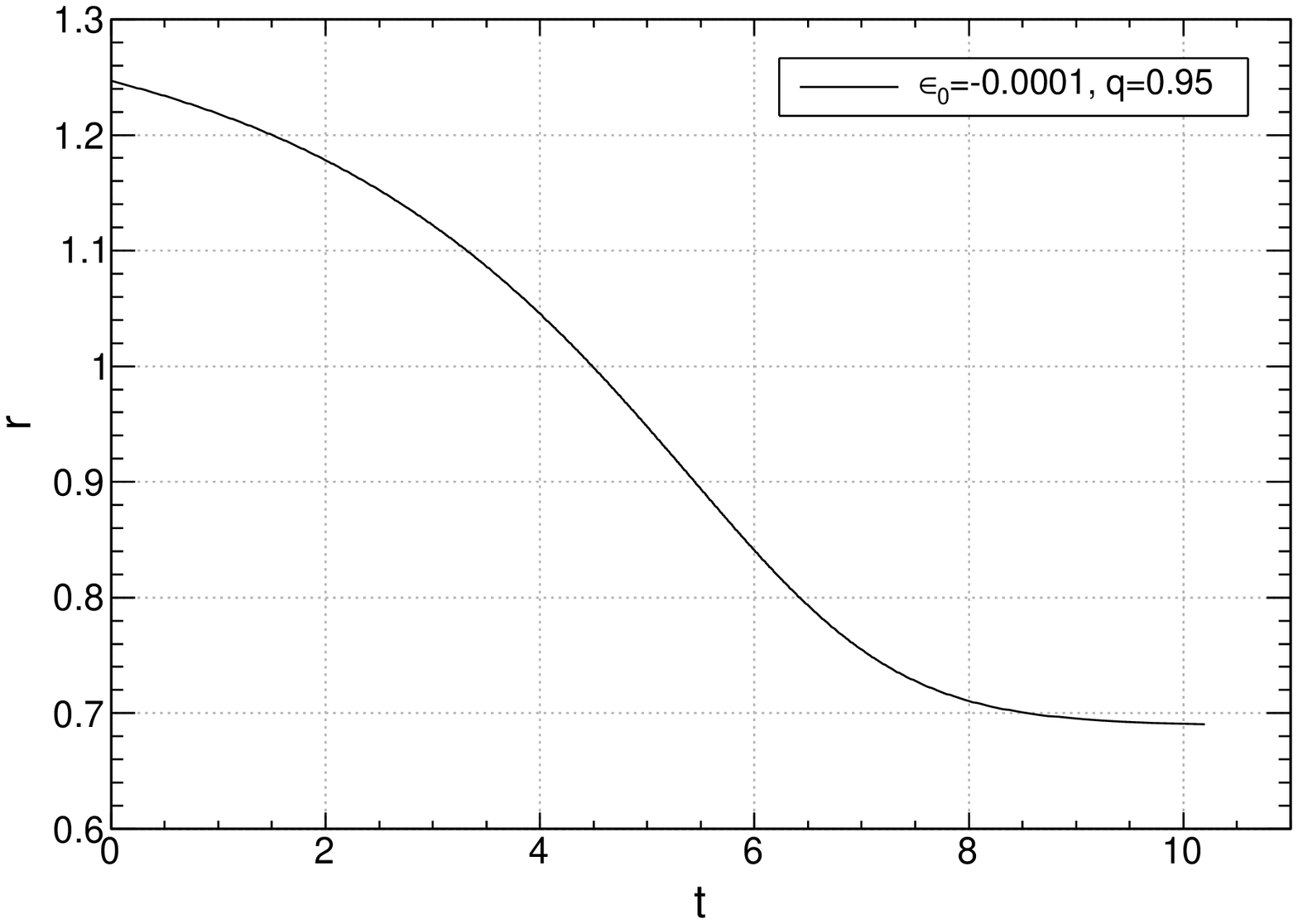}}
\subfigure[\hskip 1mm  Mass function versus $r$. \label{fig:4d}]{\includegraphics[width=0.495\textwidth]{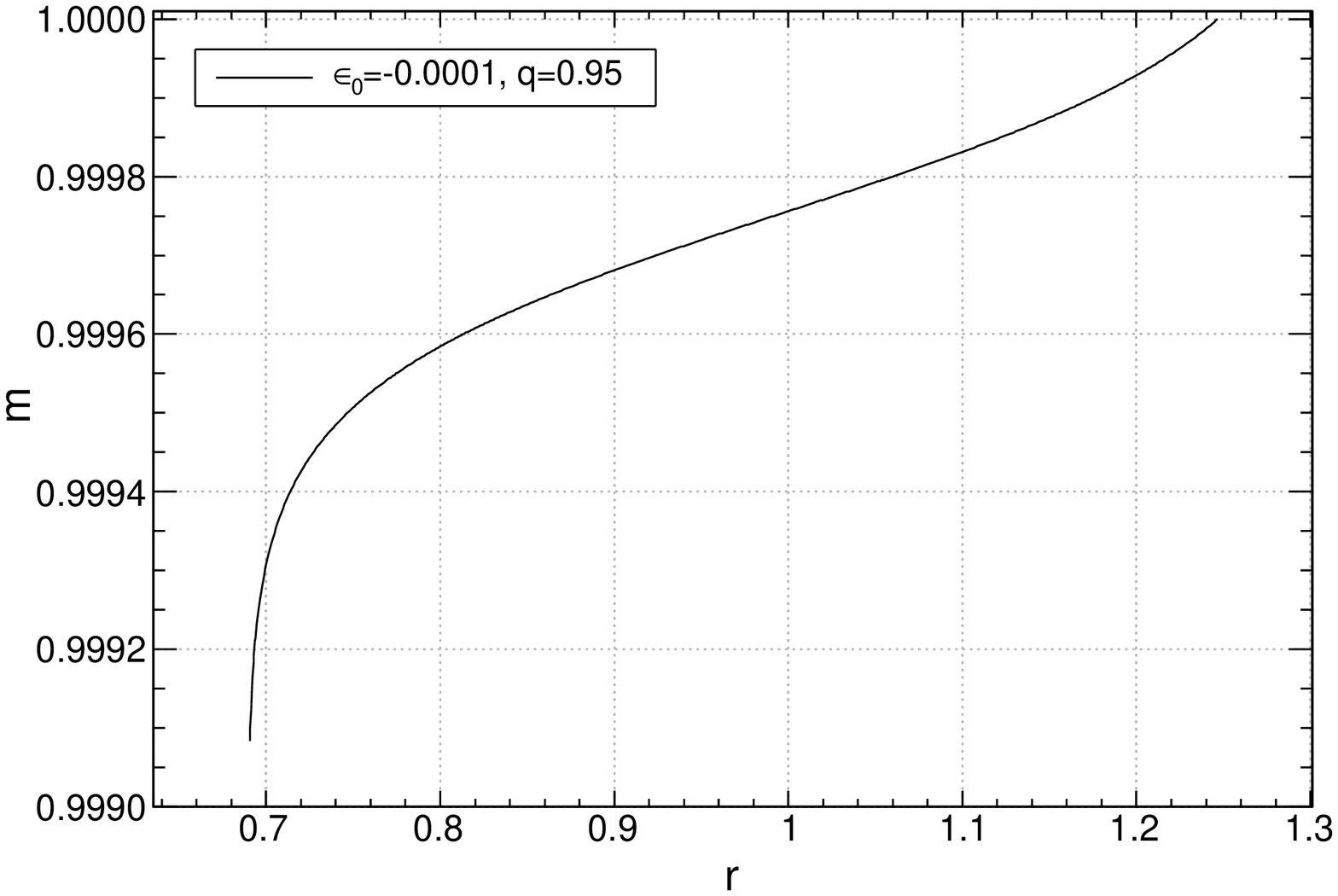}}
\caption{\label{fig:4} The case of $\epsilon_0=-0.0001, q=0.95$.}
\end{figure*} 

\begin{figure*}
\subfigure[\hskip 1mm  Metric functions $g_{tt}$ and  $|g_{rr}|$  vs
  $r$. \label{fig:5a}]{\includegraphics[width=0.495\textwidth]{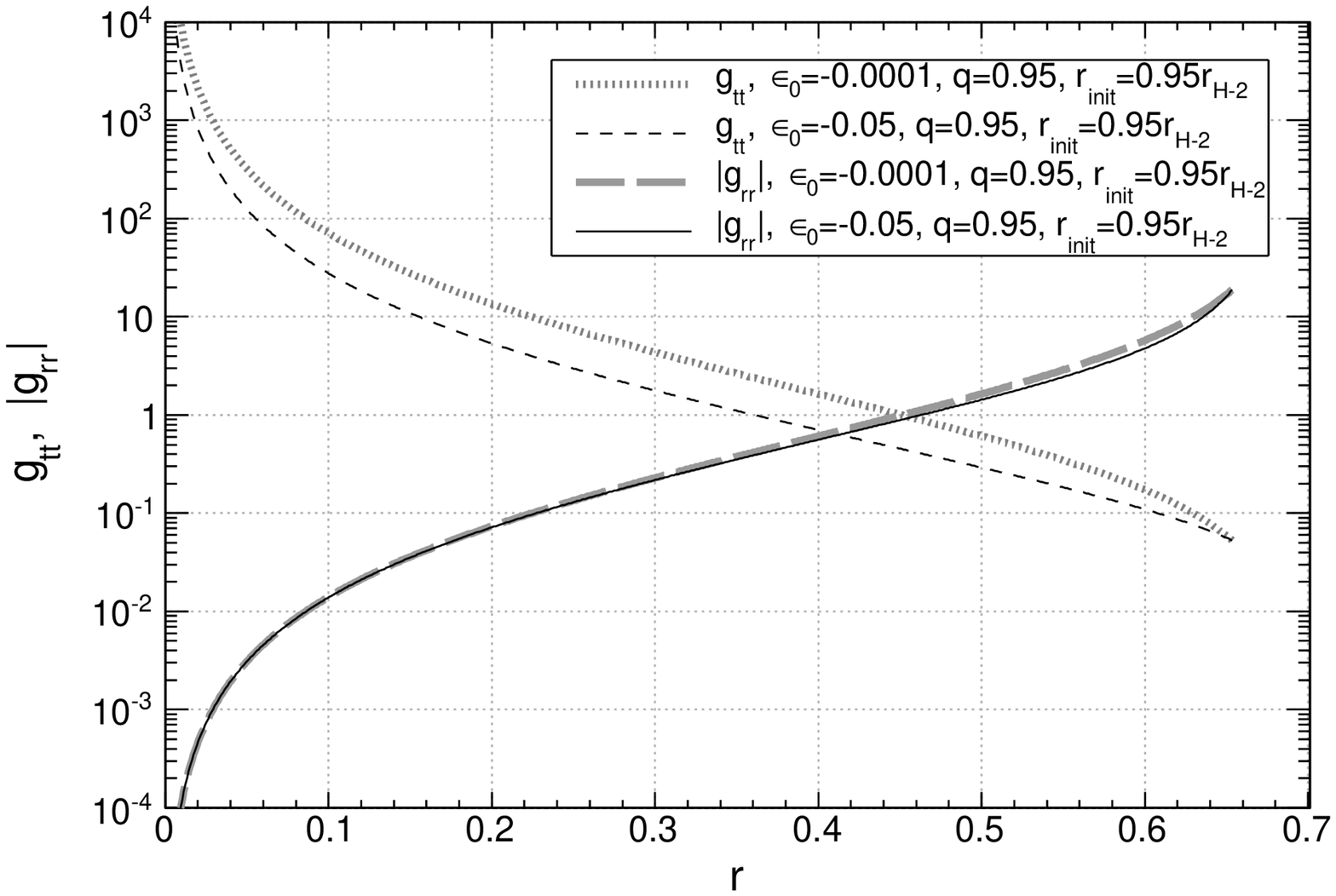}} 
\subfigure[\hskip 1mm  The $d\log g_{tt}/d\log r$ \hskip 1mm and\hskip 1mm  $d\log|g_{rr}|/d\log r$ \hskip 2mm   vs \hskip 2mm  $r$.
  \label{fig:5b}]{\includegraphics[width=0.495\textwidth]{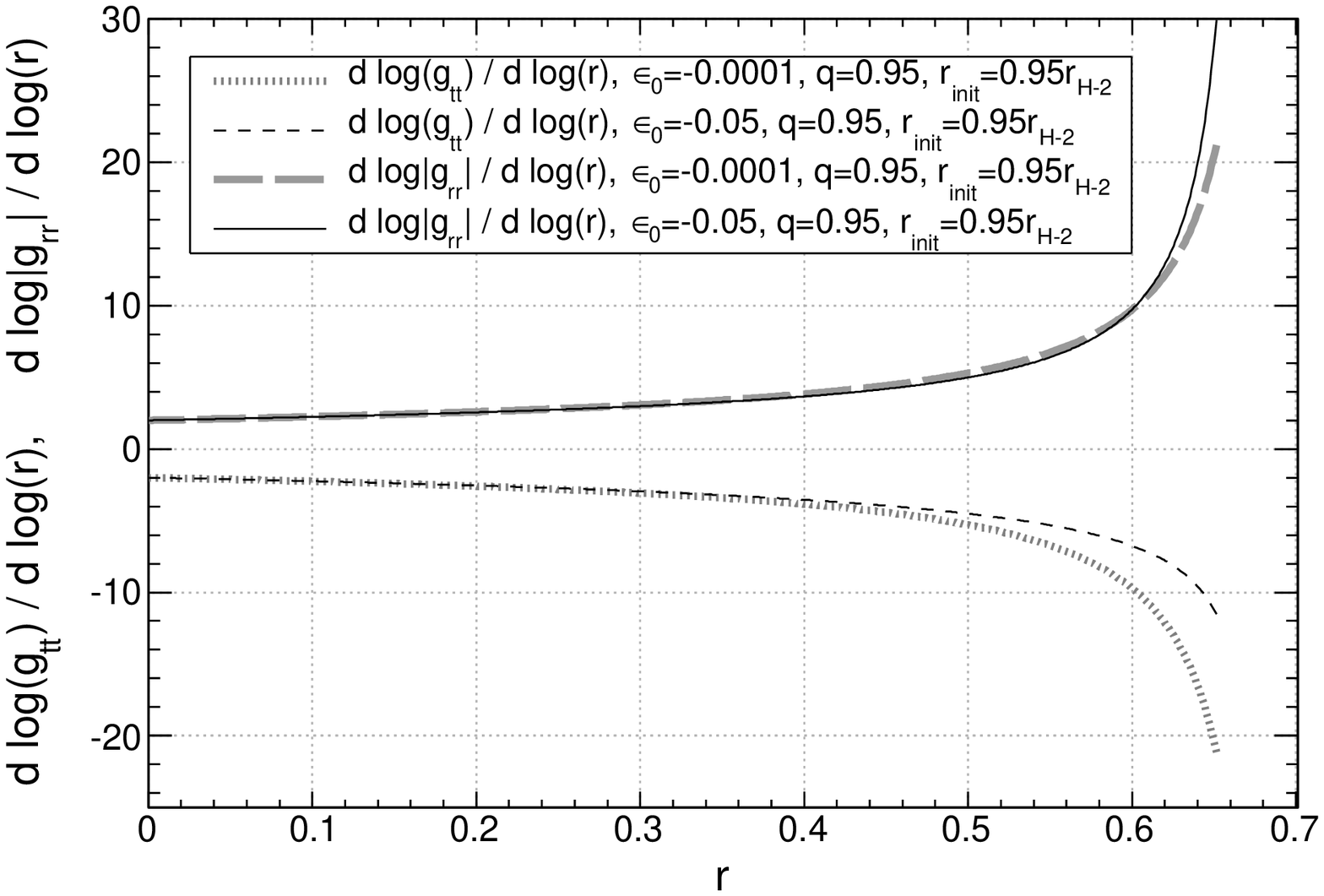}}
\subfigure[\hskip 1mm  $r$ versus $t$. \label{fig:5c}]{\includegraphics[width=0.495\textwidth]{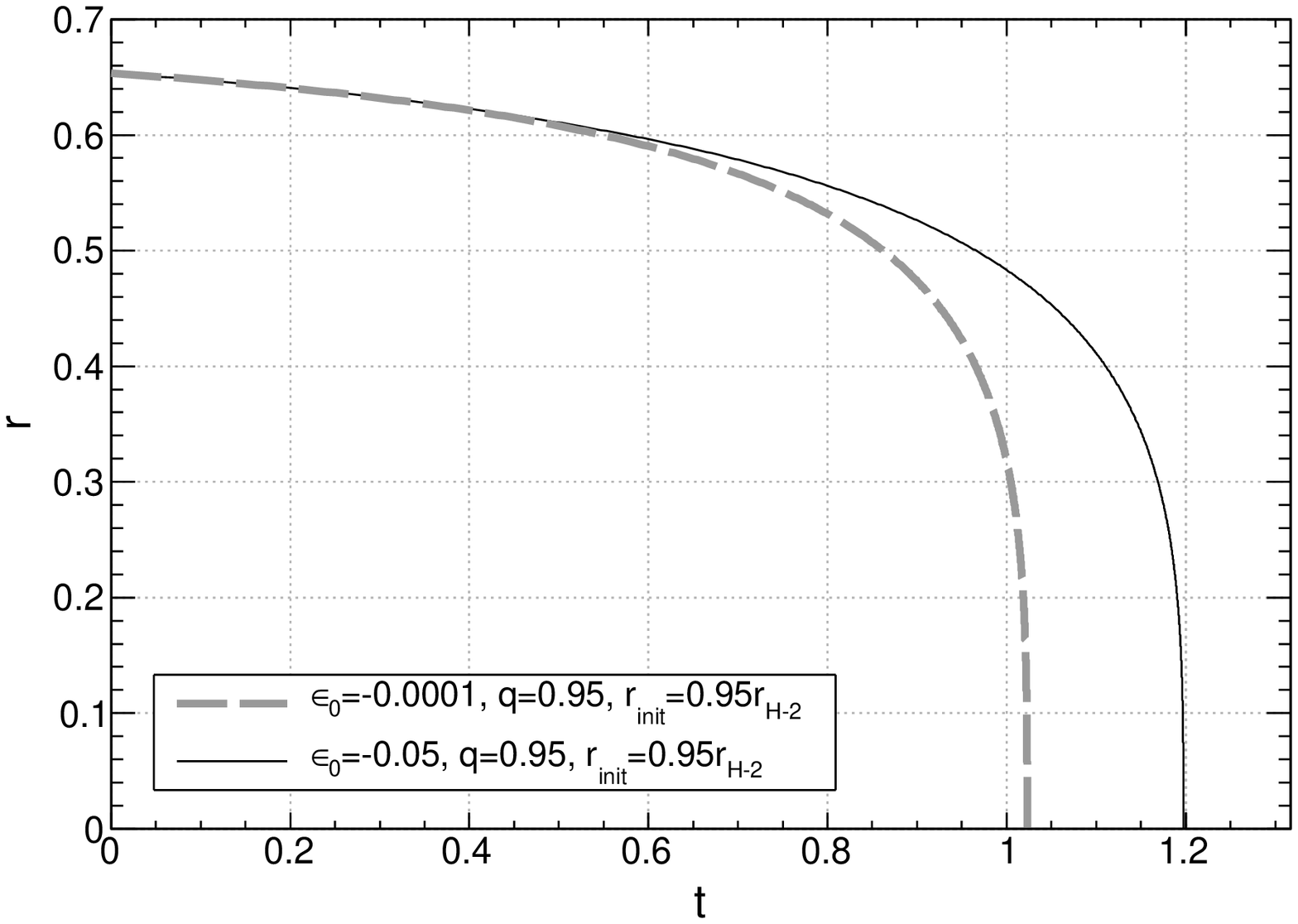}}
\subfigure[\hskip 1mm  Mass function versus $r$. \label{fig:5d}]{\includegraphics[width=0.495\textwidth]{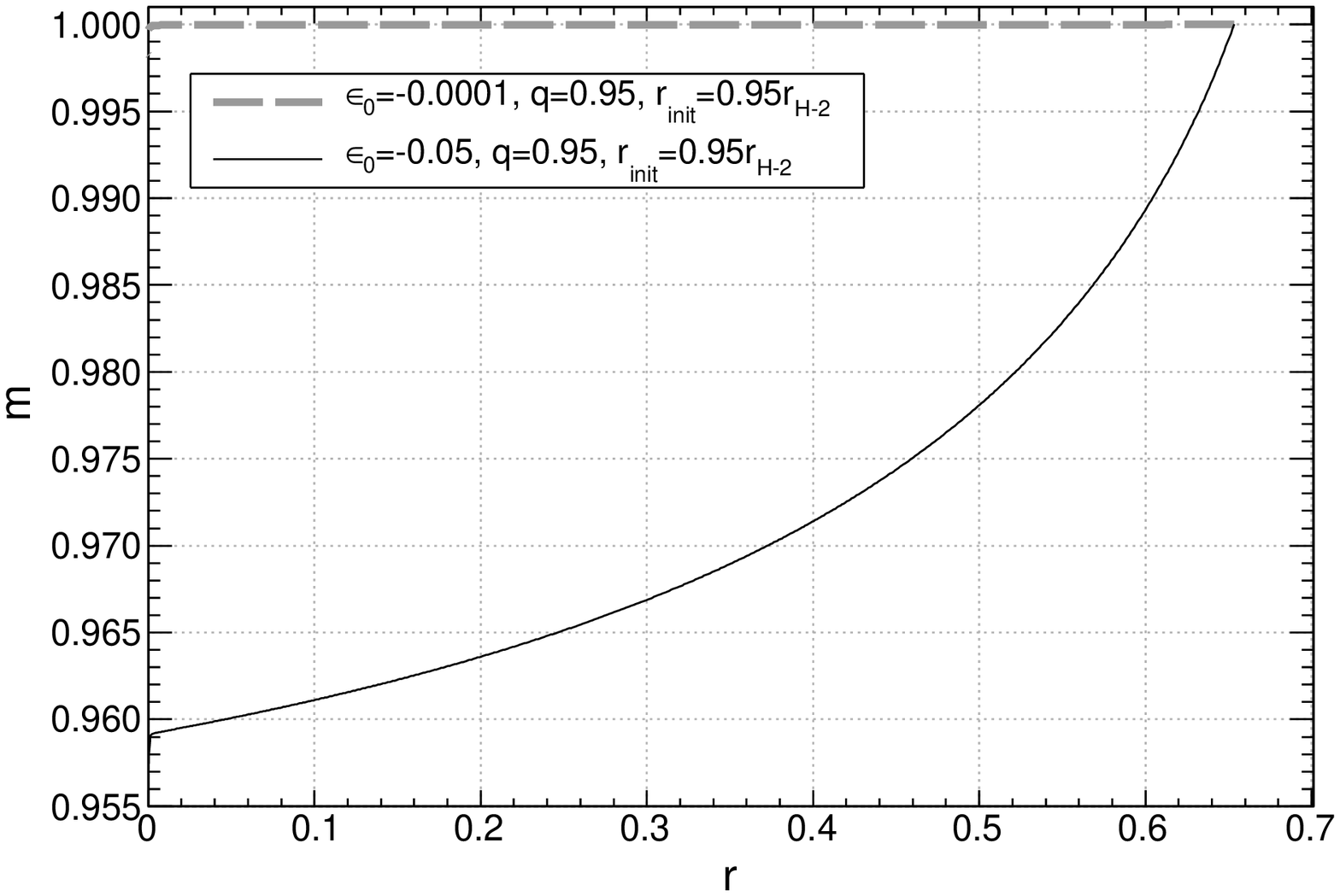}}
\caption{\label{fig:5} The case of $\epsilon_0=-0.0001, q=0.95$, $r_{init}=0.95r_{H\mbox{-}2}$,   \hskip 2mm  and $\epsilon_0=-0.05, q=0.95$, $r_{init}=0.95r_{H\mbox{-}2}$.}
\end{figure*} 

We now turn to the discussion of the case $\epsilon<0$, \hbox{$q=0$} (see
Figure \ref{fig:3}), which corresponds to the case of the collapse of the WHs analyzed numerically 
in \cite{Doroshkevich09}. Here in the limit $r\rightarrow 0$, $\beta$ is always less than \hbox{-1}. With the 
increase $\epsilon_0$  the limit $\lim_{r\rightarrow 0} |\beta|$ increases 
insignificantly. Dropping the absolute value, the limit varies from \hbox{$\beta=-1.00$} at \hbox{$\epsilon_0=-0.0001$} to \hbox{$\beta =-1.01$} 
at \hbox{$\epsilon_0=-0.05$.} The behavior of the metric functions $g_{tt}$, $g_{rr}$ is
similar for all cases $0.0001>\epsilon_0>-0.05$, see Figures  \hbox{\ref{fig:3a}--\ref{fig:3b}.}

The light-like signal comes to $r=0$ later for the larger values of $|\epsilon_0|$,
see Figure \ref{fig:3c}. 
This property is opposite to the case of $\epsilon_0>0$. The changes of the mass function 
$m$ for the cases $\epsilon_0=-0.01$ and $\epsilon_0=-0.05$ are also insignificant 
(see Figure  \ref{fig:3d}), in sharp contrast with the case $\epsilon_0>0$.

\subsection{\label{sec:model_leadOrderqNE0}Leading order analysis for the case $q\ne0$.}

We consider the leading order terms in the series expansion for the  metric functions 
and the leading order terms in the Einstein equations (\ref{eq:3}), (\ref{eq:4}), $q\neq 0$ near the 
singularity $r=0$. An assumption of $\beta >0$ makes the term with the  magnetic 
field $q^2$ to be negligible, and the analysis near $r\rightarrow 0$
becomes the same as for the cases considered in the section
\ref{sec:model_leadOrderq=0}. However, one of our conclusions in section
\ref{sec:model_leadOrderq=0} was that $\beta<-1$ (\ref{eq:17}). Thus
our current assumption of $\beta>0$ contradicts the result expressed
in (\ref{eq:17}). The case $\beta=0$ also leads to contradictions. We
come to a conclusion that the only allowed values for $\beta$ are
negative values, $\beta<0$, which  means that in the vicinity of $r=0$
we can neglect the scalar field
and consider  $d^2=0$, $q^2\neq 0$. 

To the leading order (\ref{eq:3}), (\ref{eq:4}), $d=0$ assume the forms:
\begin{equation}
 \label{eq:27}
 g\prime_{rr}r-g_{rr}-(g^2_{rr}q^2)/r^2=0,
\end{equation}
\begin{equation}
 \label{eq:28}
g\prime_{tt}r+g_{tt}+(g_{rr}g_{tt}q^2)/r^2=0. 
\end{equation}
We are looking for solutions to  (\ref{eq:27}), (\ref{eq:28}) in the vicinity $r=0$ in the form 
\begin{equation}
 \label{eq:29}
 g_{tt}^{(1)}=Br^{\beta},
\end{equation}
\begin{equation}
 \label{eq:30}
 g_{rr}^{(1)}=-Ar^{\alpha}.
\end{equation}

Substituting (\ref{eq:29}), (\ref{eq:30}) into (\ref{eq:27}), (\ref{eq:28}) we have
\begin{equation}
 \label{eq:31}
 \alpha=2, \hskip2mm \beta=-2,\hskip2mm A=-q^{-2},
\end{equation}
$B$ is arbitrary, but for the correct signature $B<0$.

Of course this is well known the Reissner-Nordstr\"{o}m solution in the vicinity of $r=0$. 
Arbitrariness of the coefficient $B$ is related to the arbitrariness of the choice of 
the $t$ coordinate. Note that now $g_{tt}<0$ and $g_{rr}>0$, this means that $t$ is 
the time coordinate, $r$ is the space coordinate, we are in the
$R$-region, see \cite {Frolov98}, and $r=0$ is a time-like singularity.

\subsection{\label{sec:model_numericalAnalysisEpsLess0}Numerical
  analysis $\epsilon<0$, $q\neq 0$.}

We start with the case $q=0.95$, $\epsilon_0=-0.0001$ and $r_{init}=0.95r_{EH}$. This case 
differs from those discussed in Section
\ref{sec:model_numericalAnalysis} by the sign of $\epsilon_0$
only. The change of sign on $\epsilon_0$ greatly influences the
physics of the processes taking place. In the case with 
$\epsilon_0>0$ the nonlinear effects of  mass inflation and focusing of the \hbox{$H\mbox{-}2$} horizon 
down to $r=0$ arized. As a result, both functions $|g_{rr}|$ and
$g_{tt}$ finally come 
to zero at $r=0$, and the space-like singularity arized (see Figure \ref{fig:1}) in the $T$-region 
(see \cite {Frolov98}). In our current case of $\epsilon<0$ there are
no mass inflation and other such effects, and 
one can observe quite different behavior (see Figure  \ref{fig:4}). The question is, can the 
collapsing configuration cross the border between the $T$-region
(as in Figure   \ref{fig:4}) and 
$R$-region and come to an internal $R$-region?  To arrive to an
answer, one needs to perform a deeper analysis and numerical calculations for the general case when $g_{rr}$ and $g_{tt}$ depend on both 
$r$ and $t$ coordinates. For now, we see that here the origin of the $r=0$ singularity 
in the $T$-region is not possible.

Thus we need to analyze here numerically the case when we are in the $R$-region and 
finally come to the time-like singularity $r=0$ as it was discussed in
Section \ref{sec:model_leadOrderqNE0}
using the leading order analysis. To accomplish that, we consider the cases(Figure  \ref{fig:5}):

\begin{eqnarray}
 \label{eq:1extra}
\epsilon_0=-0.0001, & q=0.95, & r_{init}=0.95r_{H\mbox{-}2},\\
 \label{eq:2extra}
\epsilon_0=-0.0500, & q=0.95, & r_{init}=0.95r_{H\mbox{-}2}.
\end{eqnarray}

As we emphasized in Section \ref{sec:model_leadOrderqNE0}, the asymptotic values of $g_{tt}$ and 
$g_{rr}$ do not depend on the scalar field.  Thus the dependence of the
solutions on $\epsilon_0$ is rather weak.  Note that $\alpha$ and $\beta$ come to 
their asymptotic values $\alpha=2$ and $\beta =-2$ which do not depend on $\epsilon_0$.
For larger values of  $|\epsilon_0|$ the light-like signal comes to $r=0$ at slightly larger $t$, 
and the dependence of the mass function $m$ on $r$ is stronger (but still rather weak).

\section{\label{sec:conclusions}Conclusions}

The evolution of the WHs can lead to their collapse and the origin of the singularity.
In this paper, we investigate the structure of the singularity arising as a result of the 
collapse of the WH with the exotic scalar field and the ordered magnetic field. 
We consider the spherical WHs. 

In the very vicinity of the singularity $r=0$ one can use an approximation where the 
solution depends only on one coordinate $r$ (the so-called uniform approximation). In this region, it 
is possible to consider the leading order terms in a series expansion for the metric 
functions and the leading order terms in the Einstein equations. We have demonstrated 
that in the case of the presence of the exotic scalar field only (without a magnetic 
field) the metric functions are
\begin{equation}
 \label{eq:32}
 g^{(1)}_{tt}\sim r^{\beta}, \hskip 5mm g_{rr}\sim r^{\beta+2}, \hskip 5mm  g_{22}\sim r^2, 
\end{equation}
at $r\rightarrow 0$. Here $\beta$ is a constant
\begin{equation}
 \label{eq:33}
 -2<\beta<-1. 
\end{equation}
We performed the numerical analysis of the uniform approximation to understand the 
behavior of the model as a function of its parameters. The results are described in Section\ref{sec:model_leadOrderq=0}.
For the comparison we also investigated the case of the collapse with the ordinary scalar 
field $\Phi$ with $\epsilon>0$ instead of the exotic scalar field $\Psi$ with 
$\epsilon<0$. In this case instead of inequality (\ref{eq:33}) we have
\begin{equation}
 \label{eq:34}
 \beta>-1. 
\end{equation}
and the numerical analysis shows that $\beta$ changes its sign at some critical initial
$\epsilon_0$, see Table \ref{table:1}.

There is the special interest in the case of the exotic scalar field $\Psi$ together 
with the magnetic field $H$. As it was mentioned in the Introduction (\ref{sec:intro}), the magnetic 
field in the WHs or their remnants has a special manifestation in astrophysical 
observations if the WHs really exist in the Universe.

From the leading order analysis for this case we concluded that in the
vicinity of $r=0$, 
$\beta$ is negative and the magnetic field dominates the scalar field, and the metric 
functions correspond to the Reissner-Nordstr\"{o}m solution. Thus the
singularity at $r=0$ 
is in the $R$-region and is a time-like singularity in the case of our assumptions. 
Numerical estimates demonstrate that the asymptotic behavior of the metric functions 
practically do not depend on the initial value of the exotic scalar field. We have 
investigated also the properties of the mass-function  for all cases.  
We want to emphasize the principal difference between the collapse of the ordinary 
scalar field $\Phi$ with $\epsilon_0>0$ and the magnetic field $H$, and the collapse of 
the exotic scalar field $\Psi$ with $\epsilon_0<0$ and the magnetic field $H$. 
In the case of the $\Phi$-field the collapse leads to the formation of the space-like 
singularity $r=0$ in the $T$-region, but in the case of the $\Psi$-field the 
singularity $r=0$ is a time-like and it is in the $R$-region. This is correct under the 
assumption which we have made.

At the end we want to note the following. The consideration of this
paper can be useful also for the analysis of the processes inside a BH
when it is being irradiated by an excotic scalar radiation.

\begin{acknowledgments}
The work was supported in part by Russian Foundation for Basic Research (project codes: 
08-02-00090-a, 08-02-00159-a. Scientific school 2469.2008.2 and the program Origin and 
Evolution of Stars and  Galaxies of Russian Academy of Sciences.

I.N. thanks NRL for hospitality during his visits for the work on this project.
\end{acknowledgments}

\appendix
\section{\label{append}Analysis of eqs. (\ref{eq:3}), (\ref{eq:4})}

\renewcommand{\theequation}{A.\arabic{equation}}
\setcounter{equation}{0}

We are looking for the solutions to (\ref{eq:3}), (\ref{eq:4}), $q=0$ in the vicinity $r= 0$ in 
the form
\begin{equation}
 \label{eq:A.1}
g^{(1)}_{tt}= Br^{\beta},
\end{equation}
\begin{equation}
 \label{eq:A.2}
g^{(1)}_{rr}= -Ar^{\alpha},
\end{equation}
Let us consider the case $\alpha=0$. To the leading order (\ref{eq:3}), (\ref{eq:4}), $q=0$ assume 
the form
 \begin{equation}
 \label{eq:A.3}
-A+A^2+A^2d^2B^{-1}r^{-\beta -2}=0 
\end{equation}
\begin{equation}
 \label{eq:A.4}
\beta Br^{\beta}-Br^{\beta}(-A)+Br^{\beta}-A\frac{d^2}{r^2}=0 
\end{equation}
Formula (\ref{eq:A.3}) gives
\begin{equation}
 \label{eq:A.5}
\beta =-2, 
\end{equation}
\begin{equation}
 \label{eq:A.6}
-1+A(1+\frac{d^2}{B})=0,
\end{equation}
From (\ref{eq:A.4}) and (\ref{eq:A.5}) one has 
\begin{equation}
 \label{eq:A.7}
\frac{A-1}{A}=\frac{d^2}{B} 
\end{equation}
Formulas (\ref{eq:A.6}) and (\ref{eq:A.7}) give 
\begin{equation}
 \label{eq:A.8}
A=1,  
\end{equation}
\begin{equation}
 \label{eq:A.9}
\frac{d^2}{B}=0. 
\end{equation}
Expression (\ref{eq:A.9}) contradicts the conditions $d^2\neq 0$.  Thus the case $\alpha=0$ 
is not possible.

Now let us consider the case $\alpha<0$. To the leading order Eqs (\ref{eq:3}), (\ref{eq:4}), $q=0$ assume 
the form
\begin{equation}
 \label{eq:A.10}
{g^{(1)}}^2_{rr}+{g^{(1)}}^2_{rr}\frac{d^2}{g_{tt}^{(1)} r^2}=0,
\end{equation}
\begin{equation}
 \label{eq:A.11}
-g^{(1)}_{tt}g^{(1)}_{rr}+g^{(1)}_{rr}\frac{d^2}{r^2}=0. 
\end{equation}

From (\ref{eq:A.10}) one has 
\begin{equation}
 \label{eq:A.12}
\beta=-2, 
\end{equation}
\begin{equation}
 \label{eq:A.13}
1+\frac{d^2}{13}=0. 
\end{equation}
From (\ref{eq:A.11}) and (\ref{eq:A.12}) we obtain
\begin{equation}
 \label{eq:A.14}
-1+\frac{d^2}{B}=0.
\end{equation}
Expression (\ref{eq:A.13}) contradicts (\ref{eq:A.14}). Thus the case $\alpha<0$ is not
possible.


\begin{thebibliography}{56}
\expandafter\ifx\csname natexlab\endcsname\relax\def\natexlab#1{#1}\fi
\expandafter\ifx\csname bibnamefont\endcsname\relax
  \def\bibnamefont#1{#1}\fi
\expandafter\ifx\csname bibfnamefont\endcsname\relax
  \def\bibfnamefont#1{#1}\fi
\expandafter\ifx\csname citenamefont\endcsname\relax
  \def\citenamefont#1{#1}\fi
\expandafter\ifx\csname url\endcsname\relax
  \def\url#1{\texttt{#1}}\fi
\expandafter\ifx\csname urlprefix\endcsname\relax\def\urlprefix{URL }\fi
\providecommand{\bibinfo}[2]{#2}
\providecommand{\eprint}[2][]{\url{#2}}

\bibitem[{Cambridge()}]{Carr07}
Cambridge, \emph{\bibinfo{title}{Universe or Multiverse?}}
  (\bibinfo{publisher}{Ed. by B. Carr, Cambridge Univ. Press},
  \bibinfo{year}{2007}).

\bibitem[{\citenamefont{Flamm}(1916)}]{Flamm16}
\bibinfo{author}{\bibfnamefont{L.} \bibnamefont{Flamm}},
  \bibinfo{journal}{Phys. Z.} \textbf{\bibinfo{volume}{17}},
  \bibinfo{pages}{448} (\bibinfo{year}{1916}).

\bibitem[{\citenamefont{Einstein and Rosen}(1935)}]{Einstein35}
\bibinfo{author}{\bibfnamefont{A.} \bibnamefont{Einstein}} \bibnamefont{and}
  \bibinfo{author}{\bibfnamefont{N.} \bibnamefont{Rosen}},
  \bibinfo{journal}{Phys. Rev.} \textbf{\bibinfo{volume}{48}},
  \bibinfo{pages}{73} (\bibinfo{year}{1935}).

\bibitem[{\citenamefont{Wheeler}(1955)}]{Wheeler55}
\bibinfo{author}{\bibfnamefont{J.~A.} \bibnamefont{Wheeler}},
  \bibinfo{journal}{Phys. Rev.} \textbf{\bibinfo{volume}{97}},
  \bibinfo{pages}{511} (\bibinfo{year}{1955}).

\bibitem[{\citenamefont{Wheeler}(1957)}]{Wheeler57}
\bibinfo{author}{\bibfnamefont{J.~A.} \bibnamefont{Wheeler}},
  \bibinfo{journal}{Ann. Phys.(N.Y.)} \textbf{\bibinfo{volume}{2}},
  \bibinfo{pages}{604} (\bibinfo{year}{1957}).

\bibitem[{\citenamefont{Misner and Wheeler}(1957)}]{Misner57}
\bibinfo{author}{\bibfnamefont{C.~W.} \bibnamefont{Misner}} \bibnamefont{and}
  \bibinfo{author}{\bibfnamefont{J.~A.} \bibnamefont{Wheeler}},
  \bibinfo{journal}{Ann.\ Phys.(N.Y.)} \textbf{\bibinfo{volume}{2}},
  \bibinfo{pages}{525} (\bibinfo{year}{1957}).

\bibitem[{\citenamefont{Moris and Thorne}(1988)}]{dop-0}
\bibinfo{author}{\bibfnamefont{M.~S.} \bibnamefont{Moris}} \bibnamefont{and}
  \bibinfo{author}{\bibfnamefont{K.~S.} \bibnamefont{Thorne}},
  \bibinfo{journal}{Am. J. Phys.} \textbf{\bibinfo{volume}{56}},
  \bibinfo{pages}{395} (\bibinfo{year}{1988}).

\bibitem[{\citenamefont{Visser}(1995)}]{Visser95}
\bibinfo{author}{\bibfnamefont{M.}~\bibnamefont{Visser}},
  \emph{\bibinfo{title}{Lorentzian Wormholes: from Einstein to Hawking}}
  (\bibinfo{publisher}{AIP, Woodbury}, \bibinfo{year}{1995}).

\bibitem[{\citenamefont{Ellis}(1973)}]{Ellis73}
\bibinfo{author}{\bibfnamefont{H.~G.} \bibnamefont{Ellis}},
  \bibinfo{journal}{J. Math. Phys.} \textbf{\bibinfo{volume}{14}},
  \bibinfo{pages}{104} (\bibinfo{year}{1973}).

\bibitem[{\citenamefont{Kardashev et~al.}(2006)\citenamefont{Kardashev,
  Novikov, and Shatskiy}}]{Kardashev06}
\bibinfo{author}{\bibfnamefont{N.~S.} \bibnamefont{Kardashev}},
  \bibinfo{author}{\bibfnamefont{I.~D.} \bibnamefont{Novikov}},
  \bibnamefont{and} \bibinfo{author}{\bibfnamefont{A.}~\bibnamefont{Shatskiy}},
  \bibinfo{journal}{Astron. Zh.} \textbf{\bibinfo{volume}{83}},
  \bibinfo{pages}{675} (\bibinfo{year}{2006}).

\bibitem[{\citenamefont{Kardashev et~al.}(2007)\citenamefont{Kardashev,
  Novikov, and Shatskiy}}]{Kardashev07}
\bibinfo{author}{\bibfnamefont{N.~S.} \bibnamefont{Kardashev}},
  \bibinfo{author}{\bibfnamefont{I.~D.} \bibnamefont{Novikov}},
  \bibnamefont{and} \bibinfo{author}{\bibfnamefont{A.}~\bibnamefont{Shatskiy}},
  \bibinfo{journal}{I. J. Mod. Phys. D} \textbf{\bibinfo{volume}{16}},
  \bibinfo{pages}{909} (\bibinfo{year}{2007}).

\bibitem[{\citenamefont{Shatskiy}(2007{\natexlab{a}})}]{Shatskiy07a}
\bibinfo{author}{\bibfnamefont{A.}~\bibnamefont{Shatskiy}},
  \bibinfo{journal}{Astron. Zh.} \textbf{\bibinfo{volume}{84}},
  \bibinfo{pages}{99} (\bibinfo{year}{2007}{\natexlab{a}}).

\bibitem[{\citenamefont{Morris et~al.}(1988)\citenamefont{Morris, Thorne, and
  Yurtsever}}]{Morris88}
\bibinfo{author}{\bibfnamefont{M.~S.} \bibnamefont{Morris}},
  \bibinfo{author}{\bibfnamefont{K.~S.} \bibnamefont{Thorne}},
  \bibnamefont{and}
  \bibinfo{author}{\bibfnamefont{U.}~\bibnamefont{Yurtsever}},
  \bibinfo{journal}{Phys. Rev. Lett.} \textbf{\bibinfo{volume}{61}},
  \bibinfo{pages}{1446} (\bibinfo{year}{1988}).

\bibitem[{\citenamefont{Novikov}(1989)}]{Novikov89}
\bibinfo{author}{\bibfnamefont{I.~D.} \bibnamefont{Novikov}},
  \bibinfo{journal}{JETP} \textbf{\bibinfo{volume}{95}}, \bibinfo{pages}{769}
  (\bibinfo{year}{1989}).

\bibitem[{\citenamefont{Frolov and Novikov}(1998)}]{Frolov98}
\bibinfo{author}{\bibfnamefont{V.~P.} \bibnamefont{Frolov}} \bibnamefont{and}
  \bibinfo{author}{\bibfnamefont{I.~D.} \bibnamefont{Novikov}},
  \emph{\bibinfo{title}{Black Hole Physics}} (\bibinfo{publisher}{Kluwer
  Academic Publishers}, \bibinfo{year}{1998}).

\bibitem[{\citenamefont{Thorne}(1993)}]{Thorne93}
\bibinfo{author}{\bibfnamefont{K.}~\bibnamefont{Thorne}}, in
  \emph{\bibinfo{booktitle}{GR13: General Relativity and Gravitation 1992 -�
  Proceedings of the 13th International Conference on General Relativity and
  Gravitation, Cordoba, Argentine}} (\bibinfo{publisher}{Bristol Institute of
  Physics}, \bibinfo{year}{1993}), p. \bibinfo{pages}{295}.

\bibitem[{\citenamefont{Flanagan and Wald}(1996)}]{Flanagan96}
\bibinfo{author}{\bibfnamefont{E.~E.}~\bibnamefont{Flanagan}} \bibnamefont{and}
  \bibinfo{author}{\bibfnamefont{R.~M.} \bibnamefont{Wald}},
  \bibinfo{journal}{Phys. Rev. D} \textbf{\bibinfo{volume}{54}},
  \bibinfo{pages}{6233} (\bibinfo{year}{1996}).

\bibitem[{\citenamefont{Bronnikov and Starobinsky}(2006)}]{Bronnikov06}
\bibinfo{author}{\bibfnamefont{K.~A.} \bibnamefont{Bronnikov}}
  \bibnamefont{and} \bibinfo{author}{\bibfnamefont{A.~A.}
  \bibnamefont{Starobinsky}} (\bibinfo{year}{2006}), \eprint{gr-qc/0612032}.

\bibitem[{\citenamefont{Lemos et~al.}(2003)\citenamefont{Lemos, Lobo, and
  Oliveira}}]{Lemos-Lobo}
\bibinfo{author}{\bibfnamefont{J.~P.~S.} \bibnamefont{Lemos}},
  \bibinfo{author}{\bibfnamefont{F.~S.~N.} \bibnamefont{Lobo}},
  \bibnamefont{and} \bibinfo{author}{\bibfnamefont{S.~Q.}
  \bibnamefont{deOliveira}}, \bibinfo{journal}{Phys. Rev. D}
  \textbf{\bibinfo{volume}{68}}, \bibinfo{pages}{064004}
  (\bibinfo{year}{2003}).

\bibitem[{\citenamefont{Armendariz-Picon}(2002)}]{Armendariz-Picon02}
\bibinfo{author}{\bibfnamefont{C.}~\bibnamefont{Armendariz-Picon}},
  \bibinfo{journal}{Phys. Rev. D} \textbf{\bibinfo{volume}{65}},
  \bibinfo{pages}{104010} (\bibinfo{year}{2002}).

\bibitem[{\citenamefont{Shatskiy et~al.}(2008)\citenamefont{Shatskiy, Novikov,
  and Kardashev}}]{Shatskiy08}
\bibinfo{author}{\bibfnamefont{A.}~\bibnamefont{Shatskiy}},
  \bibinfo{author}{\bibfnamefont{I.~D.} \bibnamefont{Novikov}},
  \bibnamefont{and} \bibinfo{author}{\bibfnamefont{N.~S.}
  \bibnamefont{Kardashev}}, \bibinfo{journal}{Uspekhi Fizicheskikh Nauk}
  \textbf{\bibinfo{volume}{178(5)}}, \bibinfo{pages}{481}
  (\bibinfo{year}{2008}).

\bibitem[{\citenamefont{Doroshkevich et~al.}(2008)\citenamefont{Doroshkevich,
  Kardashev, Novikov, and Novikov}}]{Doroshkevich08}
\bibinfo{author}{\bibfnamefont{A.~G.} \bibnamefont{Doroshkevich}},
  \bibinfo{author}{\bibfnamefont{N.~S.} \bibnamefont{Kardashev}},
  \bibinfo{author}{\bibfnamefont{D.~I.} \bibnamefont{Novikov}},
  \bibnamefont{and} \bibinfo{author}{\bibfnamefont{I.~D.}
  \bibnamefont{Novikov}}, \bibinfo{journal}{Astronomy Reports}
  \textbf{\bibinfo{volume}{52(8)}}, \bibinfo{pages}{616}
  (\bibinfo{year}{2008}).

\bibitem[{\citenamefont{Hayward}({\natexlab{a}})}]{dop-1}
\bibinfo{author}{\bibfnamefont{S.~A.} \bibnamefont{Hayward}},
  \eprint{gr-qc/9805019} (\bibinfo{year}{1998}).

\bibitem[{\citenamefont{Hayward}({\natexlab{b}})}]{dop-2}
\bibinfo{author}{\bibfnamefont{S.~A.} \bibnamefont{Hayward}},
  \eprint{gr-qc/0110080} (\bibinfo{year}{2001}).

\bibitem[{\citenamefont{Hayward}({\natexlab{c}})}]{dop-3}
\bibinfo{author}{\bibfnamefont{S.~A.} \bibnamefont{Hayward}},
  \eprint{gr-qc/0202059} (\bibinfo{year}{2002}).

\bibitem[{\citenamefont{Hayward and Koyama}()}]{dop-4}
\bibinfo{author}{\bibfnamefont{S.~A.} \bibnamefont{Hayward}} \bibnamefont{and}
  \bibinfo{author}{\bibfnamefont{H.}~\bibnamefont{Koyama}},
  \eprint{gr-qc/0406080} (\bibinfo{year}{2004}).

\bibitem[{\citenamefont{Koyama and Hayward}()}]{dop-5}
\bibinfo{author}{\bibfnamefont{H.}~\bibnamefont{Koyama}} \bibnamefont{and}
  \bibinfo{author}{\bibfnamefont{S.~A.} \bibnamefont{Hayward}},
  \eprint{gr-qc/0406113} (\bibinfo{year}{2004}).

\bibitem[{\citenamefont{Shatskiy}(2007{\natexlab{b}})}]{Shatskiy07b}
\bibinfo{author}{\bibfnamefont{A.}~\bibnamefont{Shatskiy}},
  \bibinfo{journal}{JETP} \textbf{\bibinfo{volume}{104}}, \bibinfo{pages}{743}
  (\bibinfo{year}{2007}{\natexlab{b}}).

\bibitem[{\citenamefont{Babichev et~al.}(2004)\citenamefont{Babichev,
  Dokuchaev, and Eroshenko}}]{Dokuchaev04}
\bibinfo{author}{\bibfnamefont{E.}~\bibnamefont{Babichev}},
  \bibinfo{author}{\bibfnamefont{V.}~\bibnamefont{Dokuchaev}},
  \bibnamefont{and}
  \bibinfo{author}{\bibfnamefont{Y.}~\bibnamefont{Eroshenko}},
  \bibinfo{journal}{Phys.Rev.Lett.} \textbf{\bibinfo{volume}{93}},
  \bibinfo{pages}{021102} (\bibinfo{year}{2004}).

\bibitem[{\citenamefont{Babichev et~al.}(2005)\citenamefont{Babichev,
  Dokuchaev, and Eroshenko}}]{Dokuchaev05}
\bibinfo{author}{\bibfnamefont{E.}~\bibnamefont{Babichev}},
  \bibinfo{author}{\bibfnamefont{V.}~\bibnamefont{Dokuchaev}},
  \bibnamefont{and}
  \bibinfo{author}{\bibfnamefont{Y.}~\bibnamefont{Eroshenko}},
  \bibinfo{journal}{JETP} \textbf{\bibinfo{volume}{100}}, \bibinfo{pages}{528}
  (\bibinfo{year}{2005}).

\bibitem[{\citenamefont{Sushkov}(2005)}]{dop-9}
\bibinfo{author}{\bibfnamefont{S.~V.} \bibnamefont{Sushkov}},
  \bibinfo{journal}{Phys. Rev. D} \textbf{\bibinfo{volume}{71}},
  \bibinfo{pages}{043520} (\bibinfo{year}{2005}).

\bibitem[{\citenamefont{Lobo}(2005{\natexlab{a}})}]{dop-10}
\bibinfo{author}{\bibfnamefont{F.~S.~N.} \bibnamefont{Lobo}},
  \bibinfo{journal}{Phys. Rev. D} \textbf{\bibinfo{volume}{71}},
  \bibinfo{pages}{084011} (\bibinfo{year}{2005}{\natexlab{a}}).

\bibitem[{\citenamefont{Lobo}(2005{\natexlab{b}})}]{dop-11}
\bibinfo{author}{\bibfnamefont{F.~S.~N.} \bibnamefont{Lobo}},
  \bibinfo{journal}{Phys. Rev. D} \textbf{\bibinfo{volume}{71}},
  \bibinfo{pages}{124022} (\bibinfo{year}{2005}{\natexlab{b}}).

\bibitem[{\citenamefont{Shinkai and Hayward}()}]{dop-6}
\bibinfo{author}{\bibfnamefont{H.~A.}~\bibnamefont{Shinkai}}, \bibnamefont{and}
  \bibinfo{author}{\bibfnamefont{S.~A.} \bibnamefont{Hayward}},
  \eprint{gr-qc/0205041}(\bibinfo{year}{2002}).

\bibitem[{\citenamefont{Shinkai}(2002)}]{Shinkai02}
\bibinfo{author}{\bibfnamefont{H.~A.} \bibnamefont{Shinkai}},
\bibnamefont{and} \bibinfo{author}{\bibfnamefont{S.~A.} \bibnamefont{Hayward}},
  \bibinfo{journal}{Phys. Rev. D} \textbf{\bibinfo{volume}{66}},
  \bibinfo{pages}{044005} (\bibinfo{year}{2002}).

\bibitem[{\citenamefont{Gonzalez et~al.}({\natexlab{a}})\citenamefont{Gonzalez,
  Guzman, and Sarbach}}]{dop-7}
\bibinfo{author}{\bibfnamefont{J.~A.} \bibnamefont{Gonzalez}},
  \bibinfo{author}{\bibfnamefont{F.~S.} \bibnamefont{Guzman}},
  \bibnamefont{and} \bibinfo{author}{\bibfnamefont{O.}~\bibnamefont{Sarbach}},
  \eprint{gr-qc/0806.1370}.

\bibitem[{\citenamefont{Gonzalez et~al.}({\natexlab{b}})\citenamefont{Gonzalez,
  Guzman, and Sarbach}}]{dop-8}
\bibinfo{author}{\bibfnamefont{J.~A.} \bibnamefont{Gonzalez}},
  \bibinfo{author}{\bibfnamefont{F.~S.} \bibnamefont{Guzman}},
  \bibnamefont{and} \bibinfo{author}{\bibfnamefont{O.}~\bibnamefont{Sarbach}},
  \eprint{gr-qc/0806.0608}.

\bibitem[{\citenamefont{Novikov}(1964{\natexlab{a}})}]{Nov64-1}
\bibinfo{author}{\bibfnamefont{I.~D.} \bibnamefont{Novikov}},
  \bibinfo{journal}{Soobshenija GAISH} \textbf{\bibinfo{volume}{132}},
  \bibinfo{pages}{3} (\bibinfo{year}{1964}{\natexlab{a}}).

\bibitem[{\citenamefont{Novikov}(1964{\natexlab{b}})}]{Nov64-2}
\bibinfo{author}{\bibfnamefont{I.~D.} \bibnamefont{Novikov}},
  \bibinfo{journal}{Soobshenija GAISH} \textbf{\bibinfo{volume}{132}},
  \bibinfo{pages}{43} (\bibinfo{year}{1964}{\natexlab{b}}).

\bibitem[{\citenamefont{Novikov}(2001)}]{Nov01}
\bibinfo{author}{\bibfnamefont{I.~D.} \bibnamefont{Novikov}},
  \bibinfo{journal}{General Relativity and Gravitation}
  \textbf{\bibinfo{volume}{33}}, \bibinfo{pages}{2259} (\bibinfo{year}{2001}).

\bibitem[{\citenamefont{Vilenkin}(1983)}]{Vilenkin83}
\bibinfo{author}{\bibfnamefont{A.} \bibnamefont{Vilenkin}},
  \bibinfo{journal}{Phys. Rev. D} \textbf{\bibinfo{volume}{27}},
  \bibinfo{pages}{2848} (\bibinfo{year}{1983}).

\bibitem[{\citenamefont{Linde}(1986)}]{Linde86}
\bibinfo{author}{\bibfnamefont{A.~D.} \bibnamefont{Linde}},
  \bibinfo{journal}{Phys. Rev. Lett. B} \textbf{\bibinfo{volume}{175}},
  \bibinfo{pages}{395} (\bibinfo{year}{1986}).

\bibitem[{\citenamefont{Hawking}(2000)}]{Hawking2000}
\bibinfo{author}{\bibfnamefont{S.~W.}~\bibnamefont{Hawking}}, in
  \emph{\bibinfo{booktitle}{Black Holes and the Structure of the
      Universe}}
 (\bibinfo{publisher}{Ed. by C.~Teitelboim and J.~Zanelli, World
    Sci.}, \bibinfo{address}{Singapore}, \bibinfo{year}{2000}), p.
  \bibinfo{pages}{23}.

\bibitem[{\citenamefont{Lobo}(2005)}]{Lobo05}
\bibinfo{author}{\bibfnamefont{F.~S.~N.} \bibnamefont{Lobo}},
  \bibinfo{journal}{Phys. Rev. D} \textbf{\bibinfo{volume}{71}},
  \bibinfo{pages}{084011} (\bibinfo{year}{2005}).

\bibitem[{\citenamefont{Hawking}(2005)}]{Hawking92}
\bibinfo{author}{\bibfnamefont{S.~W.} \bibnamefont{Hawking}},
  \bibinfo{journal}{Phys. Rev. D} \textbf{\bibinfo{volume}{46}},
  \bibinfo{pages}{603} (\bibinfo{year}{1992}).

\bibitem[{\citenamefont{Aref'eva and Volovich}()}]{Aref07}
\bibinfo{author}{\bibfnamefont{I.~Ya.} \bibnamefont{Aref'eva}} \bibnamefont{and}
  \bibinfo{author}{\bibfnamefont{I.~V.} \bibnamefont{Volovich}},
  \eprint{hep-ph/0710.2696v2}  (\bibinfo{year}{2007}).

\bibitem[{\citenamefont{Shatskiy}(2004)}]{Shatskiy04a}
\bibinfo{author}{\bibfnamefont{A.~A.} \bibnamefont{Shatskiy}},
  \bibinfo{journal}{Astron. Zh.} \textbf{\bibinfo{volume}{81}},
  \bibinfo{pages}{579} (\bibinfo{year}{2004}).

\bibitem[{\citenamefont{Shatskiy}(2004)}]{Shatskiy04b}
\bibinfo{author}{\bibfnamefont{A.~A.} \bibnamefont{Shatskiy}},
  \bibinfo{journal}{Astron. Rep.} \textbf{\bibinfo{volume}{48}},
  \bibinfo{pages}{525} (\bibinfo{year}{2004}).

\bibitem[{\citenamefont{Lobo}(2008)}]{Lobo08}
\bibinfo{author}{\bibfnamefont{F.~S.~N.}~\bibnamefont{Lobo}}
  \emph{\bibinfo{booktitle}{Classical and Quantum Gravity Research Progress}}
 (\bibinfo{publisher}{Nova Science Publisher}, \bibinfo{year}{2008}), p.
  \bibinfo{pages}{1}.

\bibitem[{\citenamefont{Cherepashchuk}(2005)}]{Cherepashchuk05}
\bibinfo{author}{\bibfnamefont{A.~M.} \bibnamefont{Cherepashchuk}},
  \bibinfo{journal}{Vestn. Mosk. Gos. Univ., Ser.~3: Fiz. Astron.} 
  \textbf{\bibinfo{volume}{No.2}},
  \bibinfo{pages}{62} (\bibinfo{year}{2005}).

\bibitem[{\citenamefont{Hayward}({\natexlab{a}})}]{Hayward09}
\bibinfo{author}{\bibfnamefont{S.~A.} \bibnamefont{Hayward}},
  \eprint{arXiv: 0903.5438} (\bibinfo{year}{2009}).

\bibitem[{\citenamefont{Hong}(2008)}]{Hong08}
\bibinfo{author}{\bibfnamefont{S.~E.} \bibnamefont{Hong}},
\bibinfo{author}{\bibfnamefont{D.~il} \bibnamefont{Hwang}},
\bibinfo{author}{\bibfnamefont{E.~D.} \bibnamefont{Stewart}} \bibnamefont{and}
\bibinfo{author}{\bibfnamefont{D.~han} \bibnamefont{Yeom}},
  \eprint{arXiv: 0808.1709} (\bibinfo{year}{2008}).

\bibitem[{\citenamefont{Yeom}(2008)}]{Yeom08}
\bibinfo{author}{\bibfnamefont{D.~han} \bibnamefont{Yeom}} \bibnamefont{and}
\bibinfo{author}{\bibfnamefont{H.} \bibnamefont{Zoe}},
  \eprint{arXiv: 0811.1637} (\bibinfo{year}{2008}).

\bibitem[{\citenamefont{Gonzalez}(2009)}]{Gonzalez09}
\bibinfo{author}{\bibfnamefont{J.~A.} \bibnamefont{Gonzalez}} \bibnamefont{and}
\bibinfo{author}{\bibfnamefont{F.~S.} \bibnamefont{Guzman}},
  \eprint{arXiv: 0903.0881} (\bibinfo{year}{2009}).

\bibitem[{\citenamefont{Doroshkevich}({\natexlab{a}})}]{Doroshkevich09}
\bibinfo{author}{\bibfnamefont{A.} \bibnamefont{Doroshkevich}},
\bibinfo{author}{\bibfnamefont{J.} \bibnamefont{Hansen}},
\bibinfo{author}{\bibfnamefont{I.} \bibnamefont{Novikov}},
\bibnamefont{and} \bibinfo{author}{\bibfnamefont{A.} \bibnamefont{Shatskiy}},
  \eprint{arXiv: 0812.0702v2} (\bibinfo{year}{2009}).

\bibitem[{\citenamefont{Burko}(1997)}]{Burko97}
\bibinfo{author}{\bibfnamefont{L.~M.}~\bibnamefont{Burko}}
\bibnamefont{and} \bibinfo{author}{\bibfnamefont{A.}
  \bibnamefont{Ori}}, \bibnamefont{eds.},
  \bibinfo{journal}{Internal structure of black holes spacetime singularities} \textbf{\bibinfo{volume}{13}},
  \emph{\bibinfo{booktitle}{of the Annals of the Israel Physical
      Society}}
\bibinfo{address}{Jerusalem}
 \bibinfo{publisher}{ISBN 0-7503-05487}, \bibinfo{year}{1997}.

\bibitem[{\citenamefont{Burko}(1998)}]{Burko98}
\bibinfo{author}{\bibfnamefont{L.~M.} \bibnamefont{Burko}},
  \bibinfo{journal}{Phys. Rev. D} \textbf{\bibinfo{volume}{58}},
  \bibinfo{pages}{084013} (\bibinfo{year}{1998}).

\bibitem[{\citenamefont{Hansen et~al.}(2005)\citenamefont{Hansen, Khokhlov, and
  Novikov}}]{Hansen05}
\bibinfo{author}{\bibfnamefont{J.}~\bibnamefont{Hansen}},
  \bibinfo{author}{\bibfnamefont{A.}~\bibnamefont{Khokhlov}}, \bibnamefont{and}
  \bibinfo{author}{\bibfnamefont{I.}~\bibnamefont{Novikov}},
  \bibinfo{journal}{Phys. Rev. D} \textbf{\bibinfo{volume}{71}},
  \bibinfo{pages}{064013} (\bibinfo{year}{2005}).


\bibitem[{\citenamefont{Doroshkevich}(1978)}]{Doroshkevich78}
\bibinfo{author}{\bibfnamefont{A.~G.} \bibnamefont{Doroshkevich}}, \bibnamefont{and}
  \bibinfo{author}{\bibfnamefont{I.~D.}~\bibnamefont{Novikov}},
  \bibinfo{journal}{Zh. Eksp. Teor. Fiz.} \textbf{\bibinfo{volume}{74}},
  \bibinfo{pages}{3} (\bibinfo{year}{1978})
 [\bibinfo{journal}{Sov. Phys. JETP} \textbf{\bibinfo{volume}{47}},
  \bibinfo{pages}{1} (\bibinfo{year}{1978})].

\end{thebibliography}
\end{document}